\documentclass[twocolumn]{aastex61}

\usepackage{color}

\def\kms{~{\rm km~s^{-1}}}
\def\cm3{~{\rm cm^{-3}}}
\def\yr{~{\rm yr}}

\def\muG{~{\mu\rm G}}

\received{}
\revised{}
\accepted{}

\submitjournal{The Astrophysical Journal}

\shorttitle{Shock Acceleration Model for Radio Relics}
\shortauthors{Kang, Ryu, \& Jones}

\begin{document}

\title{Shock Acceleration Model for the Toothbrush Radio Relic}

\author{Hyesung Kang}
\affiliation{Department of Earth Sciences, Pusan National University, Busan 46241, Korea}
\author{Dongsu Ryu}
\affiliation{Department of Physics, School of Natural Sciences, UNIST, Ulsan 44919, Korea}
\affiliation{Korea Astronomy and Space Science Institute, Daejeon 34055, Korea}
\author{T. W. Jones}
\affiliation{School of Physics and Astronomy, University of Minnesota, Minneapolis, MN 55455, USA}
\correspondingauthor{Hyesung Kang}
\email{hskang@pusan.ac.kr}

\begin{abstract}
Although many of the observed properties of giant radio relics detected in the outskirts of galaxy clusters can be 
explained by relativistic electrons accelerated at merger-driven shocks, significant puzzles remain. 
In the case of the so-called Toothbrush relic,
the shock Mach number estimated from X-ray observations ($M_{\rm X}\approx1.2-1.5$) is substantially weaker 
than that inferred from the radio spectral index ($M_{\rm rad}\approx2.8$).
Toward understanding such a discrepancy, we here consider the following diffusive shock acceleration (DSA) models:
(1) {\it weak-shock models} with $M_{\rm s}\la2$ and a preexisting population of cosmic-ray electrons (CRe) 
with a flat energy spectrum,
and (2) {\it strong-shock models} with $M_{\rm s}\approx3$ and either shock-generated suprathermal electrons 
or preexisting fossil CRe.
We calculate the synchrotron emission from the accelerated CRe,
following the time evolution of the electron DSA, and
subsequent radiative cooling and postshock turbulent acceleration (TA). 
We find that both models could reproduce reasonably well the observed integrated radio spectrum of 
the Toothbrush relic,
but the observed broad transverse profile requires the stochastic acceleration by downstream turbulence, which we label ``turbulent acceleration'' or TA to distinguish it from DSA.
Moreover, to account for the almost uniform radio spectral index profile along the length of the relic,
the weak-shock models require a preshock region over 400~kpc with a uniform population of preexisting CRe
with a high cutoff energy ($\ga40$ GeV). Due to the short cooling time, it is challenging to
explain the origin of such energetic electrons. 
Therefore, we suggest the strong-shock models with low-energy seed CRe ($\la150$~MeV)
are preferred for the radio observations of this relic.
\end{abstract}


\section{INTRODUCTION}

Some galaxy clusters contain diffuse, peripheral radio sources on scales as large as $\sim$2 Mpc in length, 
called `giant radio relics' \citep[see, e.g.,][for reviews]{feretti12,brug12, brunetti2014}. 
Typically they show highly elongated morphologies, radio spectra relatively constant along the length of the relic, 
but steepening across the width, and high linear polarization \citep{ensslin98, vanweeren10}.
Moreover, they have integrated radio spectra with a power-law form at low frequencies, but that steepen above gigaherts frequencies \citep{stroe16}. 
Previous studies have demonstrated that such observational features can often be understood as synchrotron emission 
from $\ga10$ GeV electrons in $\sim \mu$G magnetic fields, accelerated 
via diffusive shock acceleration (DSA) at merger-driven shock waves in the cluster periphery \citep[e.g.,][]{kang12}.

Yet, significant questions remain in the merger-shock DSA model of radio relics. Three of the troublesome issues are (1) low DSA efficiencies predicted for electrons injected {\it in situ} and accelerated at weak, $M_{\rm s}\lesssim 3$, shocks that are
expected to form in merging clusters \citep[e.g.,][]{ryu03, kang12}; (2) inconsistencies of the X-ray based shock 
strengths with radio synchrotron-based shock strengths with the X-ray measures typically indicating weaker shocks \citep[e.g.,][]{akamatsu13,ogrean14}; and
(3) a low fraction ($\lesssim 10$ \%) of observed merging clusters with detected radio relics \citep[e.g.,][]{ensslin01,kang16a}.
According to structure formation simulations, the mean separation 
between shock surfaces is $\sim 1$ Mpc, and the mean lifetime of intracluster medium (ICM) shocks is $t_{\rm dyn}\sim 1$ Gyr 
\citep[e.g.,][]{ryu03,pfrommer2006,skillman2008,vazza09}.
So, actively merging clusters are expected to contain several shocks, and we might actually expect multiple radio relics in typical systems. 
Some of these difficulties could be accounted for in
a scenario in which a shock may light up as a radio relic only when it 
encounters a preexisting cloud of fossil relativistic electrons in the ICM \citep[e.g.,][]{ensslin99,kangryu15}.

Here, we focus on issue (2) above. To set up what follows, we note that in the test-particle DSA model for a {\it steady, planar} shock, nonthermal electrons that are injected {\it in situ} and accelerated at
a shock of sonic Mach number $M_{\rm s}$ form a power-law momentum distribution,
$f_{\rm e}(p,r_{\rm s})\propto p^{-q}$ with $q = 4M_{\rm s}^2/(M_{\rm s}^2-1)$ \citep{dru83}.
Based on this result and the relation $\alpha_{\rm sh} = (q-3)/2$ between $q$ and the synchrotron spectral index at the shock, $\alpha_{\rm sh}$ (with $j_{\nu} \propto \nu^{-\alpha}$), the Mach number of the hypothesized relic-generating shock is then commonly inferred from its radio spectral index 
using the relation $\alpha_{\rm sh} = (M_{\rm rad}^2+3)/2(M_{\rm rad}^2-1)$.
On the other hand, the shock Mach number can be also estimated from the temperature discontinuity obtained from X-ray observations,
using the shock jump condition, $T_2/T_1=(M_{\rm X}^2+3)(5M_{\rm X}^2-1)/16/M_{\rm X}^2$, where the subscripts, $1$ and $2$, identify the upstream and downstream states, respectively. 
Sometimes, if the temperature jump is poorly constrained, $M_{\rm X}$ is assessed from an estimate of the density jump, $\sigma = \rho_2/\rho_1 = 4 M_{\rm X}^2/(M_{\rm X}^2+3)$. Although the radio and X-ray shock measures can agree, sometimes the synchrotron index, $\alpha_{\rm sh}$, implies a significantly higher Mach number, $M_{\rm rad}$, than $M_{\rm X}$.

Without subsequent, downstream acceleration, the effects of synchrotron and inverse Compton (iC) ``cooling'' ($\dot {p} \propto - p^2$) will truncate the postshock electron spectrum above energies that drop with increasing distance from the shock, since cooling times for $\ga 10$ GeV electrons under cluster conditions are generally $< 100$ Myr \citep[e.g.,][]{brunetti2014}. That is the standard explanation for observed spectral steepening across the width of the relic (downstream of the shock). For reference, we note that if the shock is steady and planar, and the postshock magnetic field is uniform, this energy loss prescription translates into an integrated synchrotron spectral index, $\alpha_{\rm int} = \alpha_{\rm sh} + 1/2$ \citep{heavens87}.

The spectral index along the northern, ``leading'' edge of
the head portion (B1) of the so-called Toothbrush relic in the merging cluster 1RXS J060303.3 at $z=0.225$ is estimated to be $\alpha_{\rm sh}\approx 0.8$ ($q \approx 4.6$) with 
the corresponding radio Mach number $M_{\rm rad}\approx 2.8$ \citep{vanweeren16}.
But the gas density jump along the same edge in B1 inferred from X-ray observations implies a much weaker shock with $M_{\rm X}\approx 1.2-1.5$ \citep{vanweeren16}. The associated radio index, $\alpha_{\rm sh} \sim 2 - 5$ ($q \sim 7 - 13$), is much too steep to account for the observed radio spectrum.

Toward understanding this discrepancy between $M_{\rm rad}$ and $M_{\rm X}$ for the Toothbrush relic, we here consider the following two scenarios for modeling the radio observations of this relic:
(1) {\it weak-shock} models ($M_{\rm s}\la 2$) with flat-spectrum, preexisting cosmic-ray elections (CRe), and
(2) {\it strong-shock} models ($M_{\rm s}\approx 3$) with low-energy seed CRe.
In the {\it weak-shock} models, 
we adopt a preshock, preexisting population of CRe with the ``right'' power-law slope, for example, 
$f_{\rm pre}(p)\propto p^{-s} \exp[-(p/p_{e,c})^2]$ with $s= 2\alpha_{\rm sh} + 3 \sim 4.4$, where $p_{e,c}/ m_{\rm e} c>10^4$ 
is an effective cutoff to the spectrum.
In the {\it strong-shock} models, on the other hand, $M_{\rm s}\approx M_{\rm rad}$ is chosen to match the observed radio 
spectral index, and low-energy seed CRe ($p/ m_{\rm e} c\sim 30$) are assumed to come from 
either the suprathermal tail population generated at the shock or a preexisting fossil population.

In the weak-shock models, the value for $p_{e,c}$ is critical, 
since the observed emissions at frequencies $\ga 100$ MHz generally come from electrons with 10 GeV or higher energies ($p/ m_{\rm e} c\ga 10^4$). 
So, if $p_{e,c}/ m_{\rm e} c\ll 10^4$, the preexisting electron population provides just ``low-energy seed electrons'' 
to the DSA process, which for $M_{\rm s} \la 2$ would still lead to $\alpha_{\rm sh} \ga 1.2$, and cannot produce the observed radio spectrum with $\alpha_{\rm sh}\approx 0.8$.
Hence, the weak-shock models with $M_{\rm s}\la 2$ can reproduce the observed spectral 
index profile of the Toothbrush relic only with $p_{e,c}/ m_{\rm e} c\sim 7-8\times 10^4$ \citep{kang16a}.
Consequently, in order to explain the fact that $M_{\rm rad}>M_{\rm X}$ for the Toothbrush or similar relics 
by the weak-shock models,
one should adopt a potentially radio-luminous, preexisting electron population with $p_{e,c}/ m_{\rm e} c\gg10^4$.
Preshock radio emission might be observable in this case, of course, unless the shock has already swept through
the region containing fossil electrons, so no preshock electrons remain.

Such a requirement for large $p_{e,c}$, however, poses a question about the origin of these preexisting CRe,
since the electrons with $p/ m_{\rm e} c\sim 8\times10^4$ in a $\muG$ field cool on a brief time scale of
 $t_{\rm rad}\sim 10$ Myr.
In the B1 portion of the Toothbrush relic, the spectral index is observed to be uniform over 400 kpc along its leading edge.
This means that in the weak-shock models the length scale of the preshock region containing preexisting CRe with 
a uniformly flat spectrum with large $p_{e,c}$ should be as long as 400 kpc, tangential to the shock surface.
Moreover, since the shock compression ratio is $\sigma \ga 2$ for $M_{\rm s} \ga 1.5$, while the observed radio width of 
the head is at least $150$ kpc, the width of this uniform preshock region should be $\ga 300$ kpc along the shock normal direction.
Considering the short cooling times for high-energy electrons,
it should be difficult to explain the origin of such a uniform cloud of preexisting CRe
by fossil CRe that were deposited in the past by an active galactic nucleus (AGN) jet, for instance, unless the effective electron dispersion speed 
across the preshock structure was $\ga 0.1$c, or there was a uniformly effective turbulent acceleration (TA) in effect across that volume.

In the strong-shock models with $M_{\rm s}\approx 3$, on the other hand, 
the challenge to account for the uniform spectrum at the relic edge becomes less severe, 
since the models require only low-energy seed CRe ($p/ m_{\rm e} c\sim 30$) that could be provided by either the shock-generated 
suprathermal electrons or preexisting fossil CRe ($p/ m_{\rm e} c\la 300$) with long cooling times ($t_{\rm rad}\ga 3.5$~Gyr).
In the latter case, the additional requirement for low-energy preexisting CRe 
to enhance the radio emission may explain why only a fraction of merger shocks can produce radio relics.
Such low-energy CRe may originate from previous episodes of shock or turbulence acceleration or AGN jets in the ICM \citep[e.g.,][]{ensslin99,pinzke13}.

According to simulations for the large-scale structure formation of the universe,
the surfaces of merger-driven shocks responsible for radio relics are expected to consist of multiple shocks
with different $M_{\rm s}$ \citep[see, e.g.,][]{skillman2008,vazza09}.
From mock X-ray and radio observations of relic shocks in numerically simulated clusters, \citet{hong15} showed that
the shock Mach numbers inferred from an X-ray temperature discontinuity tend to be lower than those from radio spectral indices.
This is because X-ray observations pick up the part of shocks with higher shock energy flux but lower $M_{\rm s}$,
while radio emissions come preferentially from the part with higher $M_{\rm s}$ and so higher electron acceleration.
In the strong-shock models, we assume that the B1 portion of the Toothbrush relic represents 
a portion of the shock surface with $M_{\rm s}\approx 3$,
extending over 400~kpc along the length of the relic.

It is important to note that the transverse width across the B1 component of the Toothbrush relic is about two times larger than that of another well-studied radio relic,
the so-callled Sausage relic in CIZA J2242.8+5301. The FWHM at 610 MHz, for example, is about 110 kpc for the Toothbrush relic \citep{vanweeren16},
while it is about 55 kpc for the Sausage relic \citep{vanweeren10}.
For the high-frequency radio emission 
from electrons with $p/m_{\rm e}c\sim 10^4$ radiatively cooled downstream from the shock,
the characteristic width of the relic behind a spherical shock is
\begin{eqnarray}
\Delta l_{\nu}
\approx 120\ {\rm kpc} \left( {u_2 \over 10^3 \kms}\right) \cdot Q \cdot
\left[{{\nu_{\rm obs}(1+z) \over {0.61{\rm GHz}}} }\right]^{-1/2},
\label{lwidth}
\end{eqnarray}
where $u_2$ is the flow speed immediately downstream of the shock and $z$ is the redshift of host clusters \citep{kang16b}.
We will argue below that relic-producing merger shocks are more spherical in geometry than planar.
Note, then, that since the downstream flow speed in the shock rest frame increases behind a spherical shock, the advection length 
in a given time scale is somewhat longer than that estimated for a planar shock \citep{donnert16}.
The factor $Q$ depends on the postshock magnetic field strength, $B_2$, as 
\begin{equation}
Q(B_2,z)\equiv \left[ { {(5\muG)^2} \over {B_2^2+B_{\rm rad}(z)^2}}\right] \left({B_2 \over 5 \muG}\right)^{1/2},
\label{qfactor}
\end{equation}
where $B_2$ and $B_{\rm rad}=3.24\muG(1+z)^2$ are expressed in units of $\mu$G. 
The factor $Q$ evaluated, for instance, for $z=0.225$ peaks with $Q_{\rm max}\approx 0.6$ with $B_2\approx 2.8\muG$. Then, with $u_2\approx 10^3\kms$,
the maximum width at 610 MHz becomes $\Delta l_{\nu}\approx 65$ kpc. 
Being only about half the observed width of the B1 region of the Toothbrush relic at this frequency, 
it seems too small to allow the width to be set by radiative cooling alone following acceleration at the shock surface. 

To overcome such a mismatch, we
here consider and include the process in which electrons are additionally accelerated stochastically by MHD/plasma turbulence
behind the shock, that is, TA.
Along somewhat similar lines, \citet{fujita2015} recently suggested that radio spectra harder than predicted by the DSA in a weak shock
could be explained if relativistic electrons are reaccelerated through resonant interactions with strong Alfv\'enic turbulence
developed downstream of the relic shock.
However, on small scales, Alfv\'enic MHD turbulence is known to become highly anisotropic, so resonant scattering is weak and ineffective at particle acceleration \citep[e.g.,][]{brunetti2007}. On the other hand, fast-mode compressive turbulence remains isotropic down to dissipation scales, 
so it has become favored in treatments of stochastic reacceleration of electrons 
producing radio halos during cluster mergers \citep{brunetti2007,brunetti2011}.
{\bf We emphasize, at the same time, that solenoidal turbulence, likely to be energetically dominant on large scales, could still play a reacceleration role through turbulent magnetic reconnection \citep[e.g.,][]{brunetti2016} or generation of small-scale slow-mode MHD waves that might interact resonantly with CRe \citep[e.g.,][]{lynn2014}.} In our study we do not depend on TA
to produce a flat electron spectrum at the shock, but rather explore its potential role as an effective means of slowing energy loss downstream of the shock.

In the next section, we describe our numerical simulations incorporating both DSA and 
TA for shock-based models designed to explore this problem.
In Section 3, our results are compared with the observations of the Toothbrush relic.
A brief summary follows in Section 4.

\section{NUMERICAL CALCULATIONS}
 
\subsection{DSA Simulations for 1D Spherical Shocks}

According to cosmological simulations \citep[e.g.,][]{ryu03,vazza09,hong14}, the formation and evolution of cluster shocks can be quite 
complex and transient with time scale $\la 1$ Gyr, but the overall morphologies of shock surfaces could be represented by 
partial surfaces of spherical bubbles blowing outward.
As in \citet{kangryu15}, we here attempt to follow for $\la 0.2$ Gyr the evolution of a 1D spherical shock, which accounts for 
deceleration and adiabatic expansion behind the shock (see the Section 2.4 for the details).

In our simulations, the diffusion-convection equation for a relativistic electron population is solved 
in 1D spherical geometry:
\begin{eqnarray}
& &{\partial g_{\rm e}\over \partial t} + u {\partial g_{\rm e} \over \partial r} \nonumber\\
& &= {1\over{3r^2}} {{\partial (r^2 u) }\over \partial r} \left( {\partial g_{\rm e}\over
\partial y} -4g_{\rm e} \right) 
+ {1 \over r^2}{\partial \over \partial r} \left[r^2 \kappa(r,p){\partial g_{\rm e} \over \partial r} \right] \nonumber\\
& &+ p {\partial \over \partial y}\left[ {D_{pp} \over p^3} \left( {\partial g_{\rm e}\over \partial y} -4g_{\rm e} \right) \right] 
+ p {\partial \over {\partial y}} \left( {b\over p^2} g_{\rm e} \right),
\label{diffcon}
\end{eqnarray}
where $g_{\rm e}(r,p,t)=f_{\rm e}(r,p,t) p^4$ is the pitch-angle-averaged phase space distribution function
of electrons, $r$ is the radial distance from the cluster center, and $y \equiv \ln(p/m_{\rm e} c)$, 
with the electron mass, $m_{\rm e}$, and the speed of light, $c$ \citep{skill75}.
The background flow velocity, $u(r,t)$, is obtained by solving the usual gas dynamic conservation equations 
in the test-particle limit where the nonthermal pressure is dynamically negligible.

The spatial diffusion coefficient for relativistic electrons is assumed to have 
the following Bohm-like form:
\begin{equation}
\kappa(r,p) = \kappa^* \left({p \over m_{\rm e} c}\right),
\label{Bohm}
\end{equation}
where $\kappa^*= k_{\rm Bohm}\cdot m_{\rm e} c^3/(3eB)$
and $k_{\rm Bohm}\ge 1$, with the limiting value representing Bohm diffusion for relativistic particles.

The electron energy loss term, $b(p)= \dot p_{\rm Coul} + \dot p_{\rm sync+iC} $, takes account of Coulomb scattering, 
synchrotron emission, and iC scattering off the cosmic microwave background (CMB) radiation \citep[e.g.,][]{sarazin99}.
For a thermal plasma with the number density $n_{\rm th}$, the Coulomb cooling rate is
 $\dot p_{\rm Coul}=3.3\times 10^{-29} n_{\rm th} [ 1 + \ln (\gamma_{\rm e}/n_{\rm th})/75]$, while the synchrotron-iC cooling rate is
$\dot p_{\rm sync+iC}=3.7\times 10^{-29} (\gamma_{\rm e}/10^4)^2 [ (B/3.24 \mu {\rm G})^2 + (1+z)^4]$ in cgs units,
where $z$ is the redshift.
Hereafter, the Lorentz factor, $\gamma_{\rm e}=p/m_{\rm e}c$, will also be used for relativistic energy.
Note that Coulomb cooling was not considered in our previous studies for DSA modeling of radio relics
\citep[e.g.,][]{kang12,kangryu15,kang16a, kang16b}.
However, since $\dot p_{\rm Coul}\gtrsim \dot p_{\rm sync+iC}$ for $\gamma_{\rm e} \lesssim 100$ in the cluster outskirts 
with $n_{\rm th}\approx 10^{-4} \cm3$, while $t_{\rm Coul} \sim p_{\rm e} /\dot p_{\rm Coul} \la$ Gyr for $\gamma_{\rm e}\la 10$, 
Coulomb cooling can, in some cases, significantly affect the electron
spectrum for $\gamma_{\rm e} \la 10^4$ and also the ensuing radio emissivity at $0.1-1$ GHz. 

The radiative cooling time due to synchrotron-iC losses is given by
\begin{equation}
t_{\rm rad} (\gamma_{\rm e}) = 9.8\times 10^{7} \yr
\left[ {(5\muG)^2} \over {B^2+B_{\rm rad}(z)^2} \right]^{2}
 \hskip-3pt\left({\gamma_{\rm e} \over 10^4 }\right)^{-1}.
\label{trad}
\end{equation}
For $B=2.5\muG$ and $z=0.225$, for example, $t_{\rm rad}\approx 8.2\times 10^7 \yr (\gamma_{\rm e}/10^4)^{-1}$.

In order to explore the effects of stochastic acceleration by turbulence, TA, we include the momentum diffusion term and implement the Crank-Nicholson scheme for it in the momentum space into the existing
CRASH numerical hydrodynamics code \citep{kj06}.
Our simulations all assume a gas adiabatic index $\gamma_g = 5/3$. 
Any nonthermal pressures from CRe and magnetic fields are dynamically insignificant in our models (see below), so they are neglected.
The physical nature of the CRe momentum diffusion coefficient $D_{pp}$ is discussed in the following section.

\subsection{Momentum Diffusion due to Turbulent Acceleration}

We pointed out in the Introduction that the recent observations of \citet{vanweeren16} showed that (1) the transverse
FWHMs of the B1 Toothbrush component are 140 kpc at 150 MHz and 110 kpc at 610 MHz, and
(2) the spectral index between the two frequencies increases from $\alpha_{150}^{610}\approx 0.8$ at the northern edge to 1.9 
at approximately 200 kpc to the south, toward the cluster center.
While the systematic spectral steepening suggests postshock electron cooling, these widths are much broader than the cooling length 
given in Equation (\ref{lwidth}). 
In effect, the spectral steepening due to radiative cooling in the postshock region is inconsistent with 
the observed profiles of radio fluxes and spectral index in this region, unless the effect of cooling is somehow substantially reduced 
(see Section 3).

In response, we explore a scenario in which the postshock electrons gain energy from turbulent waves via Fermi II acceleration, TA,
thus mitigating spectral steepening downstream. Turbulence accelerates particles stochastically; that is, if the characteristic momentum shift in a collision is $\Delta p$ and the characteristic scattering time interval is $\Delta t$, then the resulting momentum diffusion coefficient is $D_{pp} \sim (\Delta p)^2/\Delta t$. Since scattering events in turbulence typically lead to $\Delta p \propto p$, a convenient general form is 
\begin{equation}
D_{pp} = { p^2 \over {4\ \tau_{\rm acc}}}, 
\label{Dpp}
\end{equation}
where $\tau_{\rm acc} \sim (1/4) \langle(p/\Delta p)^2 \Delta t \rangle$ is an effective acceleration time scale. 
If $\tau_{\rm acc}$ is independent of momentum, this form with the factor 4 inserted into Equation (\ref{diffcon}) leads to $\tau_{\rm acc} = \langle p \rangle /(\partial \langle p \rangle /\partial t)$, 
where $\langle p \rangle$ is the mean momentum of the distribution, $f_{\rm e}(r,p,t)$.

Generally speaking, TA in an ICM context can include nonresonant scattering off compressive hydrodynamical (acoustic) turbulence \citep[e.g.,][]{ptuskin1988}, as well as gyro-resonant scattering off Alfvenic turbulence \citep[e.g.,][]{fujita2015} and Landau (also known as Cerenkov or ``transit time damping'', TTD) resonance off compressive MHD turbulence (with accompanying micro-instabilities to maintain particle isotropy; \citep[e.g.,][]{brunetti2007, brunetti2011,lynn2014}). Resonant acceleration will most often be faster than nonresonant acceleration \citep[e.g.,][]{brunetti2007,miniati2015}. Alfvenic gyro-resonance involves turbulent wavelengths comparable to particle Larmour radii, which in ICM conditions for the CRe energies of interest will be sub-astronomical unit scale. While solenoidal turbulence may very well dominate the turbulence of interest \citep[e.g.,][]{porter2015} and, in the form of Alfven waves, probably cascades to sufficiently small scales \citep[e.g.,][]{kowal2010}, it should become highly anisotropic on small scales in ICM settings and thus very inefficient in resonant scattering of CRe \citep[e.g.,][]{yan2002}\footnote{\bf We mention for completeness a proposed alternate scenario in which solenoidal turbulence leads to fast magnetic reconnection and produces a hybrid, first-second order reacceleration process \citep{brunetti2016}.}. Fast-mode, compressive MHD turbulence should remain isotropic to dissipative scales, however. 
So, even though the magnetic energy in the waves of this mode will be relatively less, they can be much more effective accelerators.

On these grounds, we adopt for our exploratory calculations a simple 
TA model based on TTD resonance with compressive, isotropic fast-mode MHD turbulence. Assuming that in the medium $\beta_p = P/P_B \gg 1$, where $P$ is the plasma thermal pressure and $P_B = B^2/(8\pi)$ is the magnetic pressure, we can then roughly express the acceleration time, $\tau_{\rm acc}$, as
\begin{equation}
\tau_{\rm acc} \sim \left(\frac{c}{a}\right)^2 \frac{1}{\langle k \rangle c} \frac{P}{W_f}.
\end{equation}
Here, $a$ is the acoustic wave speed, $W_f$ is the total energy density in fast-mode turbulence (mostly contained in compressive ``potential energy,'' but also including transverse magnetic fields essential for resonant scattering). The term $\langle k \rangle$ measures the power-spectrum-weighted mean wavenumber of the fast-mode turbulence \citep[e.g.,][]{brunetti2007}. For a power spectrum, ${\cal P}_f(k) \propto k^{-\alpha}$, over the range $2\pi/L_0\le k\le 2\pi/\ell_d$, with $3/2 \le\alpha\le 2$ \citep[e.g.,][]{brunetti2011} and $\langle 1/k \rangle =( L_0/2\pi )\mathcal{H(\alpha})$ with $\sqrt{\ell_d/L_0} \le \mathcal{H}\le 1/\ln{L_0/\ell_d}$. We can roughly estimate an outer scale, $L_0 \sim 100$ kpc behind
the shock of interest. The fast-mode dissipation scale, $\ell_d$, is uncertain and dependent on plasma collisionality, but it is likely to be less than $\sim 1$ kpc \citep[e.g.,][]{schek2006,brunetti2007,brunetti2011}. Putting these together, we can estimate $1/(\langle k\rangle c) \sim 10^4$ yrs. With an acoustic speed, $a \sim 10^3\kms $, and an estimate $W_f \sim (1/10) P$ for shock-enhanced fast-mode turbulence in the immediate postshock flow, we obtain a rough estimate of $\tau_{\rm acc,0} \sim 100$ Myr near the shock. 
As a simple model allowing for decay of this turbulence behind the shock, 
we apply the form $W_f \propto \exp\left[{{-(r_{\rm s}-r)} / r_{\rm dec}} \right]$ with $r_{\rm s}>r$,
where $r_{\rm s}$ is the radius of the spherical shock. 
So the TA time scale increases behind the shock as
\begin{equation}
\tau_{\rm acc} = \tau_{\rm acc,0} \cdot \exp\left[{{(r_{\rm s}-r)} \over r_{\rm dec}} \right]
\end{equation}
with, in most of our simulations, $r_{\rm dec}\approx 100$ kpc.

\subsection{DSA Solutions at the Shock}

Since the time scale for DSA at the shock is much shorter than the cooling time scale for radio-emitting electrons ($\sim 100$ Myr),
we assume that electrons are accelerated almost {\it instantaneously} to the maximum energy at the shock front.
On the other hand, the minimum diffusion length scale to obtain converged
solutions in simulations for Equation (\ref{diffcon}) is much smaller than the typical downstream cooling length of $\sim 100$ kpc.
Taking advantage of such disparate scales, we adopt {\it analytic} solutions for the electron spectrum at the shock location as
$f(r_{\rm s},p)= f_{\rm inj}(p)$ or $f_{\rm reacc}(p)$,
while Equation (\ref{diffcon}) is solved outside the shock.
Here, $f_{\rm inj}(p)$ represents the electrons injected {\it in situ} and accelerated at the shock, 
while $f_{\rm reacc}(p)$ represents the reaccelerated electrons preexisting in the preshock region.
So, basically we follow the energy losses and TA of electrons behind the shock,
while the DSA analytic solutions are applied to the zone containing the shock. Note that shocks in CRASH are true discontinuities and tracked on sub-grid scales \citep{kj06}.
Since we do not need to resolve the diffusive shock precursor or follow the DSA process in detail,
this scheme allows us to use a much coarser grid, reducing dramatically the required computation time.

The electron population injected {\it in situ} from the background plasma and accelerated by DSA at the shock is modeled as
\begin{equation}
f_{\rm inj}(r_{\rm s}, p)= f_N \left({p \over p_{\rm inj}}\right)^{-q} \exp \left[-\left({p \over p_{\rm eq}}\right)^2\right],
\label{finj}
\end{equation}
where $f_N$, $q$, $p_{\rm inj}$, and $p_{\rm eq}$ are the normalization factor, the standard test-particle DSA power-law slope,
the injection momentum, and the cutoff momentum, respectively.
The injection momentum roughly identifies particles with gyro-radii large enough to allow a significant fraction 
of them to recross the physical shock from downstream rather than being advected downstream 
\citep[e.g.,][]{gieseler2000, kjg02, caprioli15}.
So particles with $p>p_{\rm inj}$ are assumed to participate in the Fermi I acceleration process.
According to the hybrid simulations by \citet{caprioli14}, $p_{\rm inj}\approx (3-3.5)p_{\rm th,p}$ 
for protons at quasi-parallel shocks, where $p_{\rm th,p}=\sqrt{2m_p k T_2}$ is the proton thermal momentum
and $k$ is the Boltzmann constant.

The electron injection to the DSA Fermi I process from the thermal pool is thought to be very inefficient, 
since the momentum of thermal electrons ($p_{\rm th,e}=\sqrt{2m_{\rm e} k T_2}$) is much smaller than $p_{\rm inj}$. 
Recent particle-in-cell (PIC) simulations of quasi-perpendicular shocks by \citet{guo14}, however, showed that some of the incoming 
electrons are specularly reflected at the shock ramp and accelerated via multiple cycles of shock drift acceleration (SDA), 
resulting in a suprathermal, power-law-like tail.
Those suprathermal electrons are expected to be injected to the full Fermi I acceleration and eventually accelerated 
to highly relativistic energies.
Such a hybrid process combining specular reflection with SDA and DSA between the shock ramp and upstream waves 
is found to be effective at both quasi-perpendicular and quasi-parallel collisionless shocks \citep{park15,sundberg16}.
However, the injection momentum for electrons is not well constrained,
since the development of the full DSA power-law spectrum extending to $p/m_{\rm e}c\gg 1$ has not been established in the simulations due to severe
computational requirements for these PIC plasma simulations.
Here, we adopt a simple model in which the electron injection depends on the shock strength as
$p_{\rm inj}\approx (6.4/\sigma) m_p u_{\rm s}$, in effect resulting in $p_{\rm inj}\sim 150 p_{\rm th,e}$.
For a smaller compression ratio, the ratio, $p_{\rm inj}/m_p u_{\rm s}$, is larger, so the injection becomes less efficient.

The factor $f_N$ in Equation (\ref{finj}) depends on the suprathermal electron population with $p\sim p_{\rm inj}$ in the background plasma.
We assume that the background electrons are energized via kinetic plasma processes at the shock and
form a suprathermal tail represented by a $\kappa$ distribution of $\kappa=1.6-2.5$,
rather than a Maxwellian distribution.
The $\kappa$ distribution is well motivated in collisionless plasmas such as those in ICMs, 
where nonequilibrium interactions can easily dominate for the distribution of suprathermal particles \citep{pierrard10}. 
It has a power-law-like high-energy tail, which asymptotes to the Maxwellian distribution for large $\kappa$.
The relatively large population of suprathermal particles enhances the injection fraction compared to the Maxwellian form \citep{kang14}. 
This enhancement is larger for smaller $\kappa$. 
The injection efficiency at the shock is also less sensitive to the shock Mach number, compared to that from the Maxwellian distribution.
Note, however, that the suprathermal electron population and the injection rate do not affect significantly
the shapes of the radio-emitting electron energy spectrum and the ensuing radio synchrotron spectrum,
so the adopted models for $p_{\rm inj}$ and the $\kappa$ distribution do not influence the main conclusions of this study.

The cutoff momentum in Equation (\ref{finj}) can be estimated from the condition that the DSA acceleration rate
is equal to the synchrotron/iC loss rate:
\begin{equation}
p_{\rm eq}= \gamma_{eq} m_{\rm e} c = {m_{\rm e}^2 c^2 u_{\rm s} \over \sqrt{4e^3q/27}}
\left({B_1 \over {B_{e,1}^2 + B_{e,2}^2}}\right)^{1/2} k_{\rm Bohm}^{-1},
\label{peq}
\end{equation}
where $u_{\rm s}$ is the shock speed and
$B_{\rm e}^2 = B^2 + B_{\rm rad}(z)^2$ represents the effective magnetic field strength that accounts for
both synchrotron and iC losses \citep{kang11}.
For typical parameters with $u_{\rm s}\sim 3\times 10^3 \kms$, $B_1\sim 1 \muG$, and $k_{\rm Bohm}\sim 1$, 
the cutoff momentum becomes $p_{\rm eq}/m_{\rm e} c\sim 10^8$, but the exact value is not important, 
as long as $p_{\rm eq}/m_{\rm e} c \gg10^4$.

If there is a preexisting, upstream electron population, $f_{\rm pre}(p)$, the accelerated population at the shock is given by
\begin{equation}
f_{\rm reacc}(r_{\rm s}, p)= q \cdot p^{-q} \int_{p_{\rm inj}}^p p^{\prime q-1} f_{\rm pre} (p^\prime) dp^\prime 
\label{freacc}
\end{equation}
\citep{dru83}.
In previous studies, the DSA of preexisting CR particles is commonly referred to as ``reacceleration''
\citep[e.g.,][]{kang12,pinzke13},
so we label $f_{\rm reacc}$ as the ``DSA reaccelerated'' component.
In contrast, $f_{\rm inj}$ in Equation (\ref{finj}) represents the DSA of the background suprathermal 
particles injected {\it in situ} at the shock, so we label it as the ``DSA injected'' component.
We emphasize that our DSA reacceleration models involve {\it irreversible} acceleration of preexisting CRe, 
in contrast to the adiabatic compression models of \citet{ensslin98}. 

In our simulations, the preshock electron population is assumed to have a power-law spectrum
with exponential cutoff as follows:
\begin{equation}
f_{\rm pre}(p) = f_o \cdot p^{-s} \exp \left[ - \left({p \over p_{e,c}} \right)^2 \right],
\label{fpre}
\end{equation}
where the slope $s$ is chosen to match the observed radio spectral index.
As mentioned in the Introduction, we adopt a large cutoff Lorentz factor, 
$\gamma_{e,c}= p_{e,c}/ m_{\rm e} c=10^4-10^5$, in the weak-shock models, 
while $\gamma_{e,c}=300$ in the strong-shock models (see also Table 1).
The normalization factor, $f_o$, is arbitrary in the simulations, 
since the CR pressure is dynamically insignificant (that is, in the test-particle limit).
Yet, it would be useful to parameterize it with the ratio of the CRe to the gas pressure in the preshock region, 
$N \equiv P_{\rm CRe,1}/P_1 \propto f_o$ for a given set of $s$ and $p_{e,c}$.
In the models considered here, typically $N \sim (0.05- 0.5) \%$ matches the amplitude of observed radio flux in the Toothbrush relic.

In our DSA simulations, $p_{e,c}$ is assumed for simplicity to stay constant in the preshock region 
for the duration of the simulations ($\sim 200$ Myr).
This is probably unrealistic for high-energy electrons with $\gamma_{e}>10^4$ (see Equation (\ref{trad})), 
unless preexisting electrons are accelerated continuously in the preshock region, for instance by turbulence.

\begin{deluxetable*} {lcccccccccc} [t]
\tablecaption{Model Parameters}
\tablecolumns{11}
\tablenum{1}
\tablewidth{0pt}
\tablehead{
\colhead {Model} & \colhead{$kT_1$} & \colhead{$M_{s,i}$} & \colhead{$[M_{s,o}]^a$}& \colhead{$[kT_{2,o}]^b$} & \colhead{$B_1$} & \colhead{$[B_{2,o}]^c$} & \colhead{$s$} & \colhead{$\gamma_{e,c}$} & \colhead{$\tau_{\rm acc,0}$} & \colhead{Remarks} \\
\colhead {Name} & \colhead{(keV)} & \colhead{ } & \colhead{ } & \colhead{(keV)} & \colhead{($\muG$) } & \colhead{($\muG$) } & \colhead{ } & \colhead{ } & \colhead{(Myr)} & \colhead{ }
}
\startdata
{\bf W1.7a}  & $5.2$ & $1.7$  & $1.64$ & $8.56$ & $1.5$ & $2.7$ & $4.4$ & $10^5$ & $100$ & no injection\\
{\bf W1.7b}  & $5.2$ & $1.7$  & $1.64$ & $8.56$ & $1.5$ & $2.7$ & $4.4$ & $4\times10^4$ & $100$ & no injection \\
{\bf W1.7c}  & $5.2$ & $1.7$  & $1.64$ & $8.56$ & $1.5$ & $2.7$ & $4.4$ & $10^4$ & $100$ & no injection \\
{\bf W1.7aN} & $5.2$ & $1.7$  & $1.64$ & $8.56$ & $1.5$ & $2.7$ & $4.4$ & $8\times10^4$ & - & no injection \\
{\bf W2.0a}  & $4.3$ & $2.0$  & $1.87$ & $8.23$ & $1.5$ & $2.5$ & $4.4$ & $8\times10^4$ & $100$ & no injection \\
{\bf W2.0b}  & $4.3$ & $2.0$  & $1.87$ & $8.23$ & $1.5$ & $2.5$ & $4.4$ & $4\times10^4$ & $100$ & no injection \\
{\bf W2.0c}  & $4.3$ & $2.0$  & $1.87$ & $8.23$ & $1.5$ & $2.5$ & $4.4$ & $10^4$ & $100$ & no injection \\
{\bf W2.0d}  & $4.3$ & $2.0$  & $1.87$ & $8.23$ & $1.5$ & $2.5$ & $4.4$ & $8\times10^4$ & $50$ & no injection \\
{\bf W2.0aN} & $4.3$ & $2.0$  & $1.87$ & $8.23$ & $1.5$ & $2.5$ & $4.4$ & $8\times10^4$ & - & no injection \\
{\bf S3.6a}  & $3.0$ & $3.6$  & $3.03$ & $11.2$ & $1$   & $2.5$ & $4.6$ & $3\times 10^2$ & $100$ & $\kappa=1.6$ \\
{\bf S3.6b}  & $3.0$ & $3.6$  & $3.03$ & $11.2$ & $1$   & $2.5$ & $4.6$ & $3\times10^2$ & $100$ & seed CRe\\
{\bf S3.6c}  & $3.0$ & $3.6$  & $3.03$ & $11.2$ & $1$   & $2.5$ & $4.6$ & $3\times10^2$ & $100$ & no decay ($r_{dec} \rightarrow\infty$) \\
{\bf S3.6aN} & $3.0$ & $3.6$  & $3.03$ & $11.2$ & $1$   & $2.5$ & $4.6$ & $3\times10^2$ & - & $\kappa=1.6$ \\
{\bf S3.6bN} & $3.0$ & $3.6$  & $3.03$ & $11.2$ & $1$   & $2.5$ & $4.6$ & $3\times10^2$ & - & seed CRe\\
\enddata
\tablenotetext{a}{Shock sonic Mach number at the time of observation.}
\tablenotetext{b}{Postshock temperature at the time of observation.}
\tablenotetext{c}{Postshock magnetic field strength at the time of observation.}
\vspace{-0.8cm}
\end{deluxetable*}

\subsection{Model Parameters}

\subsubsection{Observed Properties of the Toothbrush Relic}

Before outlining our simulation model parameters, we briefly review our target, the Toothbrush radio relic. The relic has a linear morphology aligned roughly east-west with multiple components that, together, resemble the head and handle
of a toothbrush \citep{vanweeren12} on respectively the west and east ends. Our focus is on the head component (labeled as B1 in Figure 4 of \citet{vanweeren12}), whose ``bristles'' point southward and whose northern edge seems to coincide with the shock location detected in X-ray
observations.
\citet{vanweeren16} estimated rather similar preshock and postshock temperatures,
$kT_1=8.3_{-2.4}^{+3.2}$ keV and $kT_2=8.2_{-0.9}^{+0.7}$ keV, respectively,
indicating that $kT_1$ is more uncertain from their data. From the slope change in the X-ray surface brightness across the putative shock in the component B1, they estimated a low shock Mach number, $M_{\rm X}\sim 1.2$. %
On the other hand, $M_{\rm rad}\approx 2.8$ is required to explain the radio spectral index ($\alpha_{\rm s}\approx 0.8$) at the northern edge of B1 as a consequence of the DSA of CRe electrons injected locally from the thermal plasma. 

\subsubsection{Shock Dynamics}

We assume for simplicity, but one step beyond a planar shock model, that the shock dynamics can be approximated initially by a self-similar blast wave
that propagates through an isothermal ICM with the density profile of $n_{\rm th}= 10^{-4} \cm3 (r/0.8 {\rm Mpc})^{-2}$.
Then, the shock radius and velocity evolve roughly as $r_{\rm s}\propto t^{2/3}$ and $u_{\rm s}\propto t^{-1/3}$,
respectively, where $t$ is the time since the nominal point explosion for the spherical blast wave \citep[e.g.,][]{ryu91}.
The shock Mach number decreases in time as the spherical shock expands in the simulations.
For this self-similar shock, the downstream flow speed {\it in the upstream rest frame} decreases toward the cluster center
as $u(r)\propto (r/r_{\rm s})$,
so the postshock flow speed with respect to the shock front increases downstream away from the shock.
We acknowledge that the actual shock dynamics in the simulations deviate slightly from such behaviors, since the model shocks are not strong, although this should not influence our conclusions.

Table 1 summarizes model parameters for the DSA simulations considered in this study. 
Considering the observed ranges for both $kT_1$ and $kT_2$, we vary 
the preshock temperature as $kT_1=3.0-5.2$ keV.
At the onset of the simulations, the shock is specified by the initial Mach number, $M_{s,i}=1.7-3.6$,
which sets the initial shock speed as $u_{\rm s,i}= M_{\rm s,i}\cdot 150\kms \sqrt{T_1/10^6{\rm K}} $,
and is located at $r_{\rm s,i}\approx 0.8$ Mpc from the cluster center. 
This can be regarded as the time when the relic-generating shock encounters the preshock region containing preexisting electrons,
that is, the birth of the radio relic.

We define the ``shock age," $t_{\rm age}\equiv t - t_{\rm onset}$, as the time since the onset of our simulations.
We find that the downstream radio flux profiles and the integrated spectrum become compatible with the observations 
at the ``time of observation," $t_{\rm age}\sim 140-150$ Myr, typically when the shock is located at $r_{\rm s}\approx 1.1-1.2$ Mpc. 
The fourth and fifth columns of Table 1 show the shock Mach number, $M_{\rm s,o}$, and the postshock temperature, 
$kT_{\rm 2,o}$, at the time of observation.

In this study, we examine if the various proposed DSA-based models can explain the observed radio flux profiles reported by \citet{vanweeren16}, 
which, as we pointed out, depend strongly on the electron cooling length behind the shock. Therefore the magnetic field strength, which impacts electron cooling, is another key parameter.
The sixth column of Table 1 shows the preshock magnetic field strength, $B_1=1-1.5 \muG$,
which is assumed to be uniform in the upstream region.
The postshock magnetic field strength is modeled as $B_2(t)=B_1 \sqrt{1/3+2\sigma(t)^2/3}\approx 2.5-2.7\muG$,
which decreases slightly as the shock compression ratio, $\sigma(t)$, decreases in time in response to shock evolution.
For the downstream region ($r<r_{\rm s}$), we assume a simple model in which the magnetic field strength scales with the 
gas pressure as $B_{\rm dn}(r,t)= B_2(t) \cdot [P(r,t)/P_{2}(t)]^{1/2}$,
where $P_{2}(t)$ is the gas pressure immediately behind the shock.

\subsubsection{DSA Model Parameters}

\begin{figure*}[t]
\vspace{-0.4cm}
\hspace{-0.2cm}
\includegraphics[scale=0.90]{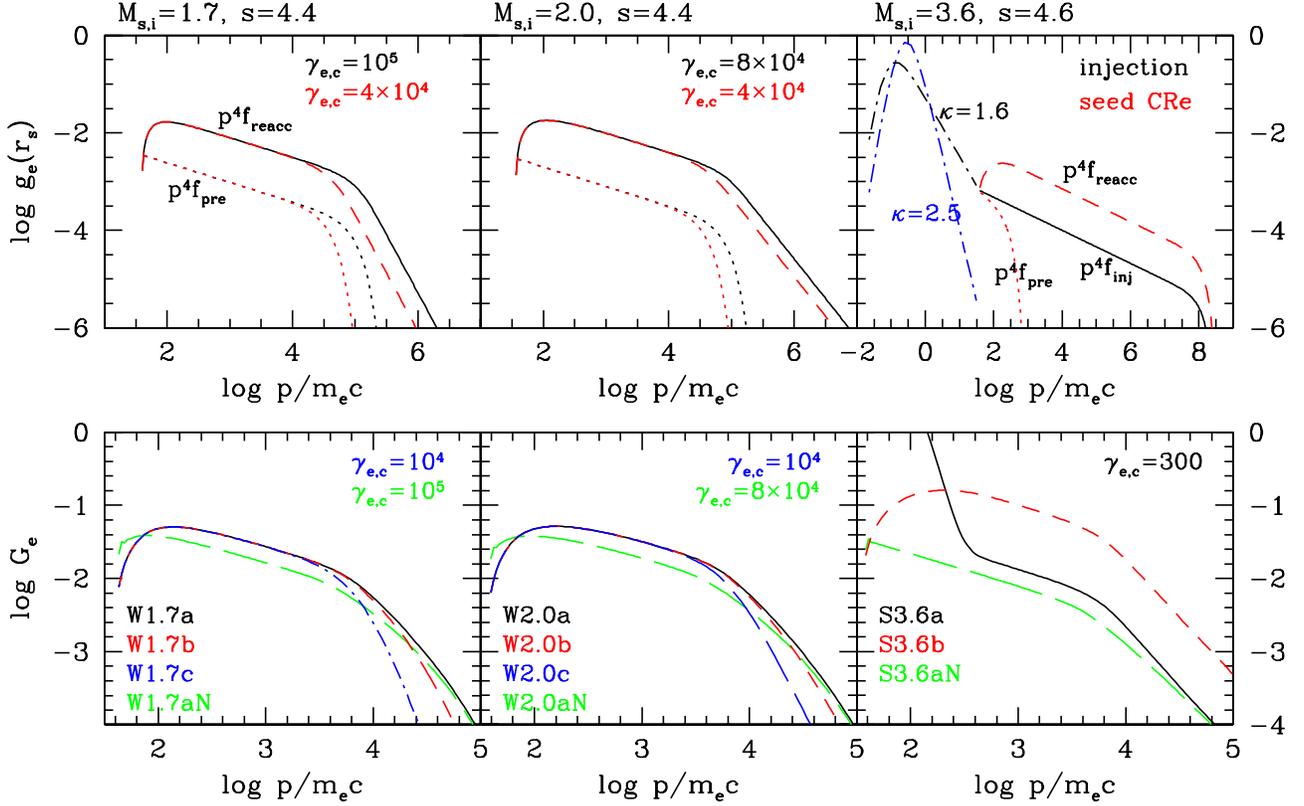}
\vspace{-7.0cm}
\caption{
Electron distribution at the shock position, $g_{\rm e}(r_{\rm s},p)=p^4 f_{\rm e}(r_{\rm s},p)$ (upper panels),
and volume-integrated electron distribution, $G_{\rm e}(p)=\int g_{\rm e}(r,p)dV$ (lower panels). See Table 1 for model parameters.
In the upper panels, the red and black dotted lines show the distribution function for preexisting electrons, $p^4f_{\rm pre}$, 
while the black solid and red dashed lines show either $p^4 f_{\rm reacc}$ for the W1.7, W2.0, and S3.6b models or  $p^4 f_{\rm inj}$ for the S3.6a model.
In the upper right panel, the $\kappa$ distributions with $\kappa=1.6$ (black dot-dashed line) and 
$\kappa=2.5$ (blue dot-dashed line) for suprathermal electrons are also shown for $p<p_{\rm inj}\approx 30 m_ec$.
In the lower panels, results are shown at $t_{\rm age}=$ 142 Myr for W1.7a (black solid lines), W1.7b (red dashed), W1.7c (blue dot-dashed) 
and W1.7aN (green long-dashed);
at $t_{\rm age}=$ 148 Myr for W2.0a (black solid), W2.0b (red dashed), W2.0c (blue dot-dashed) 
and W2.0aN (green long-dashed);
and at $t_{\rm age}=$ 144 Myr for S3.6a (black solid), S3.6b (red dashed) and S3.6aN (green long-dashed).
}
\label{Fig1}
\end{figure*}

\begin{figure*}[t]
\vspace{-1cm}
\hspace{-0.2cm}
\includegraphics[scale=0.9]{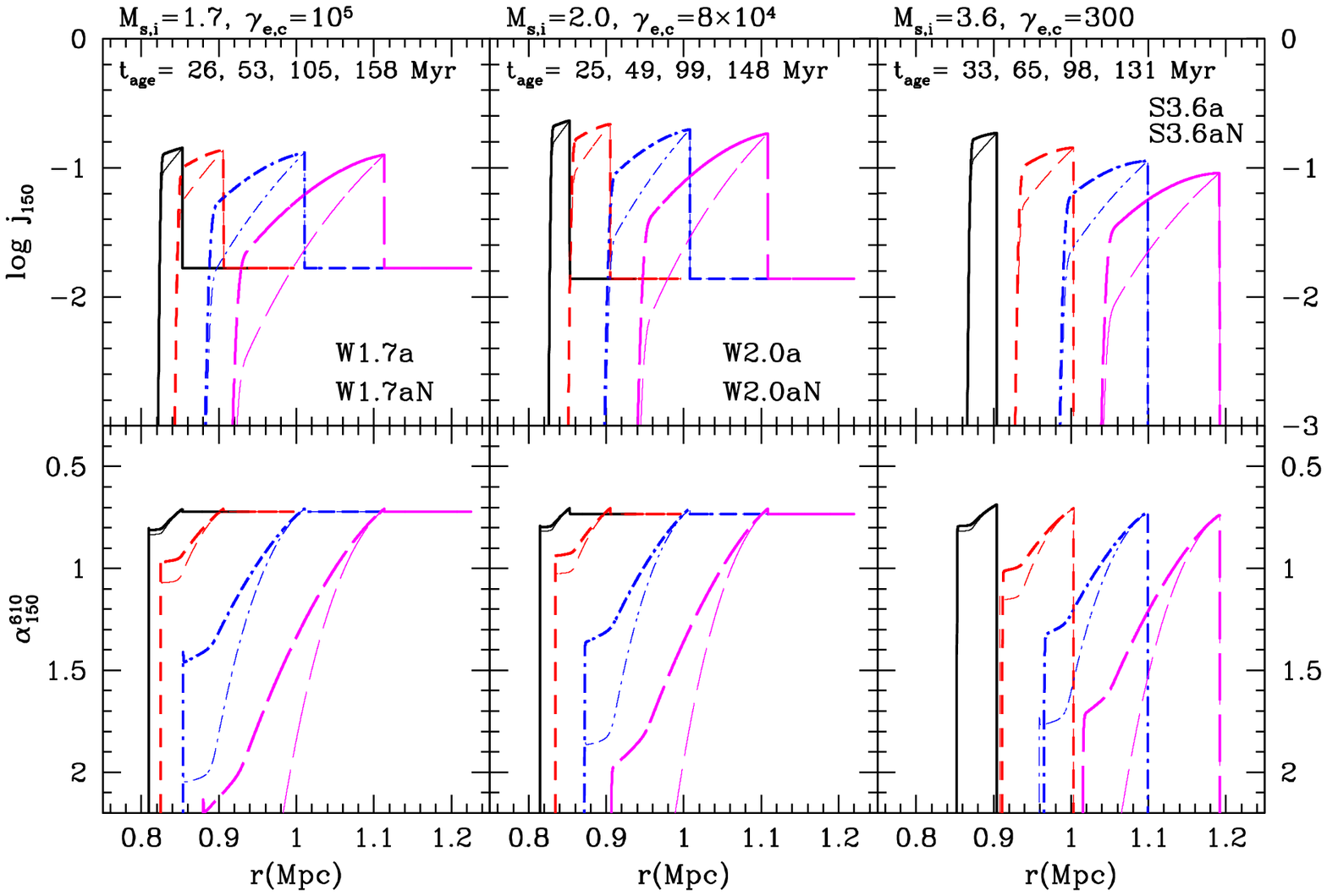}
\vspace{-6.0cm}
\caption{
Synchrotron emissivity at 150 MHz, $j_{150}(r)$ (upper panels, in arbitrary units), 
and associated spectral index between 150 and 610 MHz, $\alpha_{150}^{610}$ (lower panels),
as a function of the radial distance from the cluster center at four different $t_{\rm age}$.
See Table 1 for model parameters. 
Thick (thin) lines are used for the models with (without) turbulent acceleration.}
\label{Fig2}
\end{figure*}

As mentioned in the Introduction, 
the discrepancy between the observationally inferred values of $M_{\rm X}$ and $M_{\rm rad}$ could be resolved
if we adopt a preexisting electron population with the ``right values'' of $s$ and $p_{e,c}$.
Alternatively, we can explain the observed radio spectral index with a shock with $M_{\rm rad}$, 
assuming that $M_{\rm X}$ and $M_{\rm rad}$ may represent different parts of nonuniform shock surfaces.
Our study considers both of these possible scenarios: 
(1) in the {\it  weak-shock} models a shock with $M_{\rm s} \la 2$ encounters a preshock region of 
a flat preexisting CRe population with $\gamma_{e,c}>10^4$,
and (2) in the {\it strong-shock} models a shock with $M_{\rm s}\approx 3.0$ accelerates low-energy seed
electrons ($\gamma_{\rm e}\sim 30$), either shock-generated suprathermal electrons
or preexisting fossil CRe.

In general, we find in these experiments that the models in which postshock electrons cool without turbulent reacceleration cannot explain the 
broad widths of the observed radio flux profiles, independent of the assumed shock strength and CRe sources, as shown in the next section.
Consequently, we also explore models that include postshock TA with the characteristic acceleration time scale of
$\tau_{\rm acc} \sim100$ Myr, which, as argued in Section 2.2, is justifiable in this context and also is comparable to expected postshock electron cooling times.

To facilitate the analyses below, we comment briefly on the model naming convention in Table 1.
The first character, W or S, refers to weak-shock or strong-shock models, respectively,
while the number after the first letter corresponds to the initial Mach number, $M_{s,i}$.
This is followed by a sequence label (a, b, c, d) as the preexisting CRe cutoff, $\gamma_{e,c}$, 
or TA time, $\tau_{\rm acc}$, parameters vary. 
If there is no postshock TA, the letter ``N'' is appended at the end.
 
In the weak-shock models, we adopt the initial shock Mach number, $M_{s,i}=1.7-2.0$, and
set $s=4.4$ as the power-law slope for preexisting CRe.
In order to see the dependence of emissions on the cutoff energy in the preexisting electron spectrum, 
we consider a wide range of $\gamma_{e,c}= 10^4-10^5$ 
in the W1.7a, b, c and W2.0a, b, c models (column 9 of Table 1).
In model W2.0d, an enhanced, postshock turbulent reacceleration with shorter $\tau_{\rm acc,0}$ is considered.
For all W1.7 and W2.0 models, the {\it in situ} injection from the background plasma is turned off 
in order to focus on the ``DSA reacceleration'' of preexisting CRe.

In the case of the strong-shock scenario,
the S3.6a model includes only the ``DSA injection'' from a suprathermal 
$\kappa$ distribution of $\kappa=1.6$,
while the S3.6b model incorporates only the ``DSA reacceleration'' of 
the preexisting CRe population with $s=4.6$ and $\gamma_{e,c}= 300$.
For the S3.6b model, the simulation results remain similar for different values of cutoff energy, $\gamma_{e,c}$, as long as 
$\gamma_{e,c}>p_{\rm inj}/m_{\rm e}c\approx 30$.
In the S3.6c model, the decay of turbulence is turned off ($r_{dec}\rightarrow\infty$), 
so the momentum diffusion coefficient is assumed to be uniform behind the shock; that is,
$D_{pp} = p^2/ (4\tau_{\rm acc,0})$.

The upper panels of Figure 1 show the preexisting electron spectrum, $f_{\rm pre}$ (red and black dotted lines), and the
{\it analytic solutions} for the shock spectra, $f_{\rm reacc}$ and $f_{\rm inj}$, given in Equations (\ref{freacc}) and (\ref{finj}), respectively.
Here, the normalization for $f_{\rm pre}$ corresponds to $N \simeq 0.01$ for W1.7a and W2.0a and $N \simeq 0.001$ for S3.6b.
For the W1.7 and W2.0 models, at the shock $g_{\rm e}(r_{\rm s},p)= p^4 f_{\rm reacc}(p)$ is used, 
since the {\it in situ} injection from the background plasma is suppressed. 
For these models, the slope of $f_{\rm reacc}(p)$ at the shock position is the preshock, $s$, for $p<p_{\rm e,c}$, 
while it becomes the DSA value, $q$, for $p> p_{\rm e,c}$.

In the upper right panel of Figure 1,
the black dot-dashed line illustrates the $\kappa$ distribution of $\kappa=1.6$ for $p<p_{\rm inj}$,
while the black solid and red long-dashed lines show $f_{\rm inj}(p)$ and $f_{\rm reacc}(p)$, respectively,
for $p\ge p_{\rm inj}$.
As shown here, the normalization factor $f_N$ for $f_{\rm inj}(p)$ is specified by the $\kappa$ distribution.
In all S3.6 models, the DSA slope $q$ is flatter than $s$,
so both $f_{\rm inj}$ and $f_{\rm reacc}$ have power-law spectra with the slope $q$,
extending to $p_{\rm eq}/m_{\rm e} c \sim 10^8$, independent of $\gamma_{e,c}$.
As a result, preexisting low-energy CRe just provide seeds to the DSA process and enhance the injection,
but do not affect the shape of the postshock electron spectrum for $p\gg p_{\rm inj}$  
(i.e., the black solid line for S3.6a versus the red dashed line for S3.6b in the upper right panel). 

Regarding the shock-generated suprathermal electron population and its posited non-Maxwellian, 
$\kappa$-distribution form, the $\kappa$ index is not universal, since it depends on a local balance of 
nonequilibrium processes. If we adopt a steeper $\kappa$ distribution, with,
for example, $\kappa=2.5$ (blue dot-dashed line in Figure 1), 
then the amplitude of the injected electron flux at $p_{\rm inj}$ will be smaller, and
so the ensuing radio flux will be reduced from the models shown here (S3.6a and S3.6aN).

\begin{figure*}[t]
\vspace{-0.9cm}
\hspace{-0.2cm}
\includegraphics[scale=0.90]{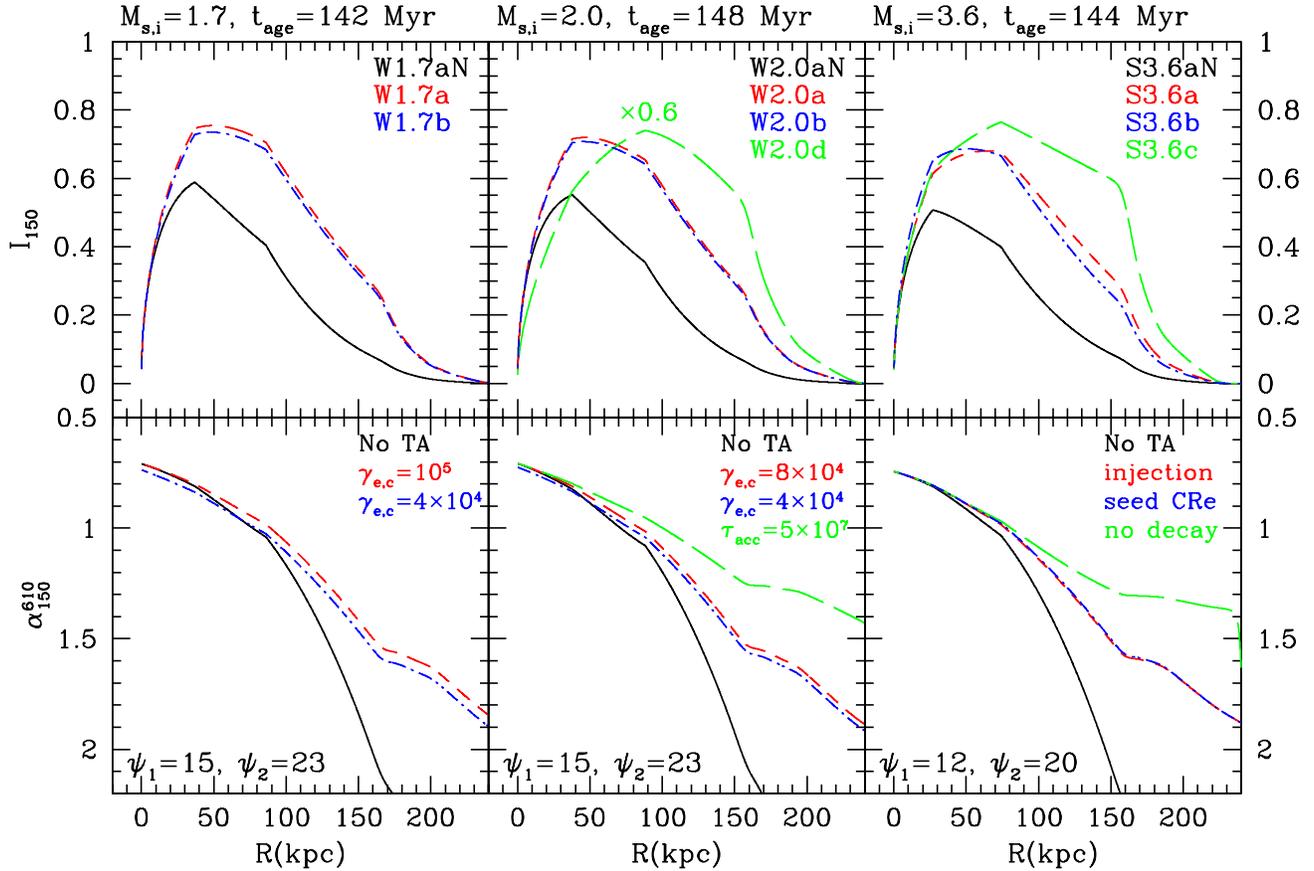}
\vspace{-6.0cm}
\caption{Surface brightness profile at 150 MHz, $I_{150}$ (upper panels, in arbitrary units),
and the spectral index between 150 and 610 MHz with $I$ (lower panels),
as a function of the projected distance behind the shock, $R$ (kpc).
See Table 1 for model parameters.
Results are shown at $t_{\rm age}=$ 142 Myr for W1.7aN (black solid lines), W1.7a (red dashed), and W1.7b (blue dot-dashed);
at $t_{\rm age}=$ 148 Myr for W2.0aN (black solid), W2.0a (red dashed), W2.0b (blue dot-dashed), and W2.0d (green long-dashed);
and at $t_{\rm age}=$ 144 Myr for S3.6aN (black solid), S3.6a (red dashed), S3.6b (blue dot-dashed), and S3.6c (green long-dashed).
The extension angles are assumed to be $\psi_1=15^{\circ}$ and $\psi_2=23^{\circ}$ for the W1.7 and W2.0 models,
while $\psi_1=12^{\circ}$ and $\psi_2=20^{\circ}$ for the S3.6 models.
The $I_{150}$ of the W2.0d model (faster TA) is reduced by a factor of 0.6,
compared to those of other W2.0 models.
}
\label{Fig3}
\end{figure*}

\section{RESULTS OF DSA SIMULATIONS}

\subsection{Radial Profiles of Radio Emissivity}

Figure 2 shows the evolution of the synchrotron volume emissivity at 150 MHz, $j_{150}(r)$, 
and the associated spectral index between 150 and 610 MHz, $\alpha_{150}^{610}(r)$, determined from
$j_{150}(r)$ and $j_{610}(r)$.
The shock is located at $r_{s,i}\approx 0.8$ Mpc at the start of the simulations, $t_{\rm age}=0$.
In the case of the W1.7 and W2.0 models, this can be regarded as the moment when the shock begins to accelerate preexisting 
electrons and become radio-bright.
The figure shows that in the models with postshock TA (thick lines) the spectral steepening is significantly delayed
relative to the models without TA (thin lines).
Only the models with TA seem to produce $\alpha_{150}^{610}$ profiles broad enough to be 
compatible with the observed profile, which increases from $\alpha_{150}^{610} \approx 0.8$ to $\alpha_{150}^{610} \approx 2.0$ over $\sim200$ kpc across the relic width.
For the W1.7a and W2.0a models, the emissivity increases by an order of magnitude (a factor $8-12$) from upstream to downstream across the shock.
Note that the subsequent, postshock emissivity decreases faster with time in the S3.6a model 
with only DSA injection from the background plasma,
compared to the W1.7 and W2.0 models with the DSA reacceleration of the preexisting CRe.
This is because for the particular injection model adopted here, 
the injection rate depends on $u_{\rm s}$ and 
$M_{\rm s}$, both of which decrease in time as the shock propagates. 

\subsection{Radio Surface Brightness Profiles}

\begin{figure*}[t]
\vspace{-0.9cm}
\hspace{-0.2cm}
\includegraphics[scale=0.9]{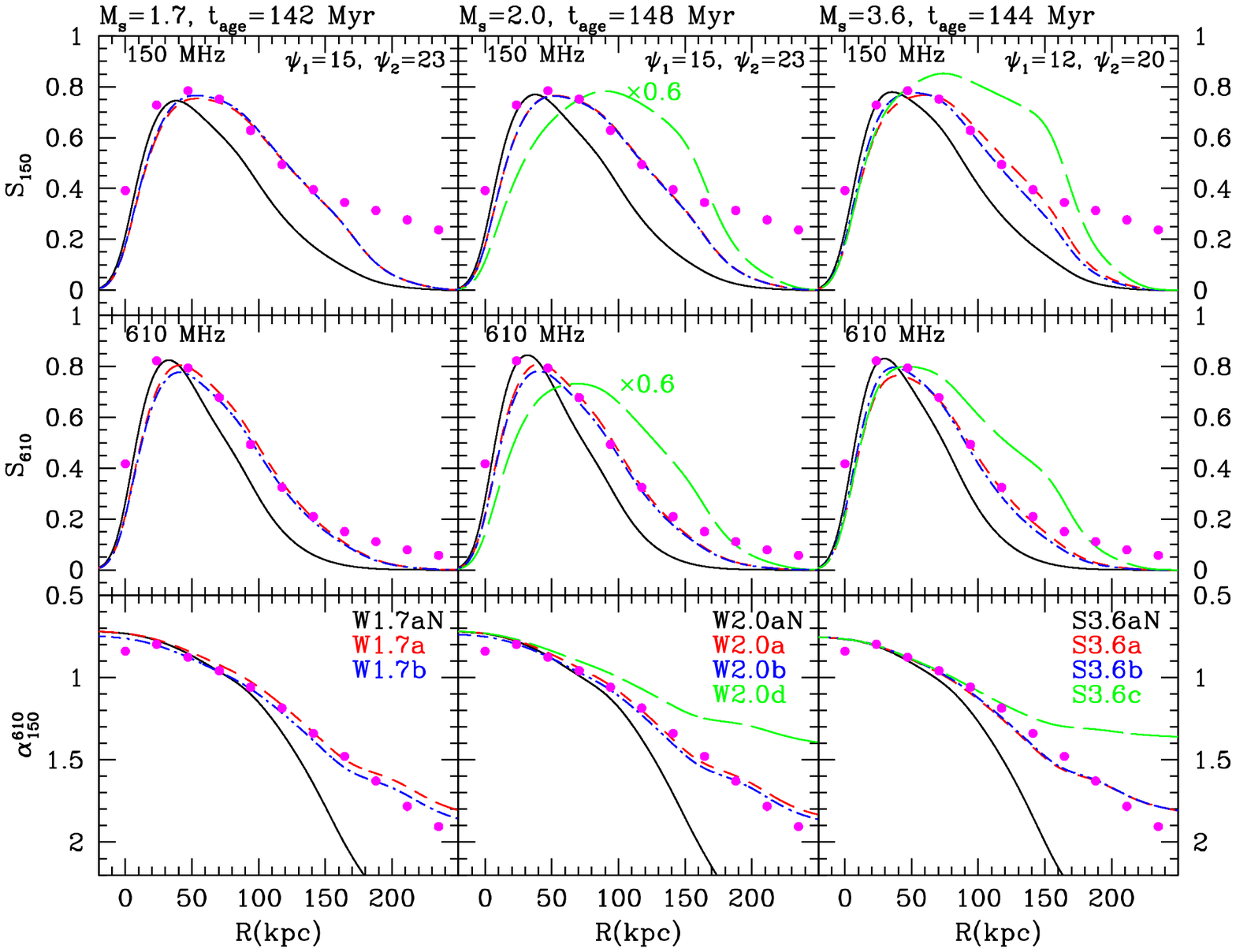}
\vspace{-4.0cm}
\caption{Radio flux density, $S_{\nu}$, within a synthesized telescope beam at 150 MHz (top panels) and at 610 MHz (middle panels)
in arbitrary units, and the
spectral index, $\alpha_{150}^{610}$, between the two frequencies (bottom panels),
plotted as a function of the projected distance behind the shock, $R$ (kpc).
See Table 1 for model parameters.
The surface brightness profiles shown in Figure 3 are smoothed by a
Gaussian beam with $6.5^{\prime\prime}$ resolution ($\approx 23.5$ kpc). 
The same line types as in Figure 3 are used.
$S_{150}$ and $S_{610}$ of the W2.0d model (faster TA) are lowered by a factor of 0.6, compared to those of the other W2.0 models, as in Figure 3.
The magenta dots are the observational data of \citet{vanweeren16}.}
\label{Fig4}
\end{figure*}

The radio surface brightness, 
$I_{\nu}$, is calculated by adopting the spherical wedge volume of radio-emitting electrons, specified with the
two extension angles relative to the sky plane, $\psi_1$ and $\psi_2$, as shown in Figure 2 of \citet{kang16a}:
\begin{equation}
I_{\nu}(R)= \int_0^{h_{\rm 1, max}} j_{\nu}(r) d {\it h_1}+ \int_0^{h_{\rm 2, max}} j_{\nu}(r) d {\it h_2},
\label{SB}
\end{equation}
where $R$ is the distance behind the projected shock edge in the plane of the sky (measured from the shock toward the cluster center),
$r$ is the radial distance outward from the cluster center,
and $h_1 = r \sin{\psi_1}$ and $h_2 = r \sin{\psi_2}$ are the path lengths along line of sight beyond and in front of the sky plane, respectively. 
(See Figure 1 of \citet{kang15} for the geometrical meaning of $R$.)

Figure 3 shows the profiles of $I_{150} (R)$ and $\alpha_{150}^{610} (R)$, now calculated from $I_{150} (R)$ and $I_{610} (R)$,
at the shock age of $t_{\rm age}=142-148$ Myr. 
The adopted values of $\psi_1$ and $\psi_2$ are given in the lower panels.
In the weak-shock models with $M_{s,o}\approx 1.6-1.9$, a high-cutoff Lorentz factor, 
$\gamma_{e,c}\ga 4\times 10^4$, is required to match $\alpha_{150}^{610}\approx 0.8$ at the shock position.
From the geometric consideration only (that is, the line-of-sight length through the model relic), 
the first inflection point in the $I (R)$ profile occurs at
$r_{\rm s}(1-\cos \psi_1)\approx 38$ kpc for the shock radius $r_{\rm s}\approx 1.1$ Mpc and $\psi_1=15^{\circ}$, and
the second inflection point occurs at $r_{\rm s}(1-\cos \psi_2)\approx 87$ kpc for $\psi_2=23^{\circ}$.
The third inflection point at $d\approx 150-160$ kpc occurs at the postshock advection length, $\sim u_2 t_{\rm age}$,
which corresponds to the width of the postshock spherical shell.

Note that the normalization factor for $I_{150}$ is arbitrary, but it is the same for all three models with $M_{s,i}=1.7$ 
(upper left panel) and for the three models with $M_{s,i}=2.0$ (upper middle panel).
But note that for the S2.0d model $I_{150}$ is reduced by a factor of 0.6, compared to the other three models.
So, for example, the relative ratio of $I_{150}$ between W1.7aN (without TA) and
W1.7a (with TA) is meaningful.
In the case of the S3.6 models (upper right panel), on the other hand, 
the normalization factor is the same for S3.6aN, S3.6a, and S3.6c 
(with only DSA injection of shock-generated suprathermal electrons),
but a different factor is used for S3.6b (with preexisting, seed CRe) in order to plot the four models together in the same panel. 

The effects of postshock TA can be seen clearly in the spectral steepening 
of $\alpha_{150}^{610}$ in the lower panels.
As shown in Figure 4 below, for instance, the S3.6aN model (black) produces a ``too-steep'' spectral profile compared to observations,
while the S3.6c model (green) without turbulence decay ($r_{dec} \rightarrow\infty$) produces a ``too-flat'' spectral profile.

To compare to the observed radio flux density distribution, $S_{\nu}$, the intensity, $I_{\nu}$, should be convolved with telescope beams.
In Figure 4, a Gaussian smoothing with 23.5 kpc width 
(equivalent to $6.\arcsec5$ at the distance of the Toothbrush relic) 
is applied to calculate $S_{\nu} (R)$,
while the spectral index $\alpha_{150}^{610}$ is then calculated from $S_{150} (R)$ and $S_{610} (R)$.
The observational data of \citet{vanweeren16} are shown with magenta dots.
The observed flux density at 150 MHz covering the region of $6.\arcsec5 \times 70\arcsec$ 
at $R\approx 50$ kpc behind the shock is $S_{150}\approx 0.20$ Jy.
The required amount of preexisting CRe to match this flux level corresponds to $N\approx 0.4-0.5 \%$ 
for the W1.7a,b and W2.0a,b models, and $N\approx0.05\%$ for the S3.6b model.
In the S3.6a model (without preexisting CRe), the corresponding flux density is $S_{150}\approx 0.004$ Jy,
five times smaller than the observed value.
Considering that the $\kappa=1.6$ distribution is already quite flat and so $\kappa$ index cannot be 
reduced further, it could be difficult to increase significantly the flux density $S_{150}$ in the S3.6a model.
In that regard, the S3.6b model with preexisting CRe is favored over the S3.6a model.
Note that the synchrotron intensity scales with $I_{150}\propto B_2^{(s-1)/2}$, 
while the downstream magnetic field strength in these models is chosen to be $B_{2,o}\approx 2.5-2.7\muG$ 
(see Table 1) in order to maximize the downstream cooling length given in Equation (\ref{lwidth}).

\begin{figure*}[t]
\vspace{-0.8cm}
\hspace{-0.2cm}
\includegraphics[scale=0.9]{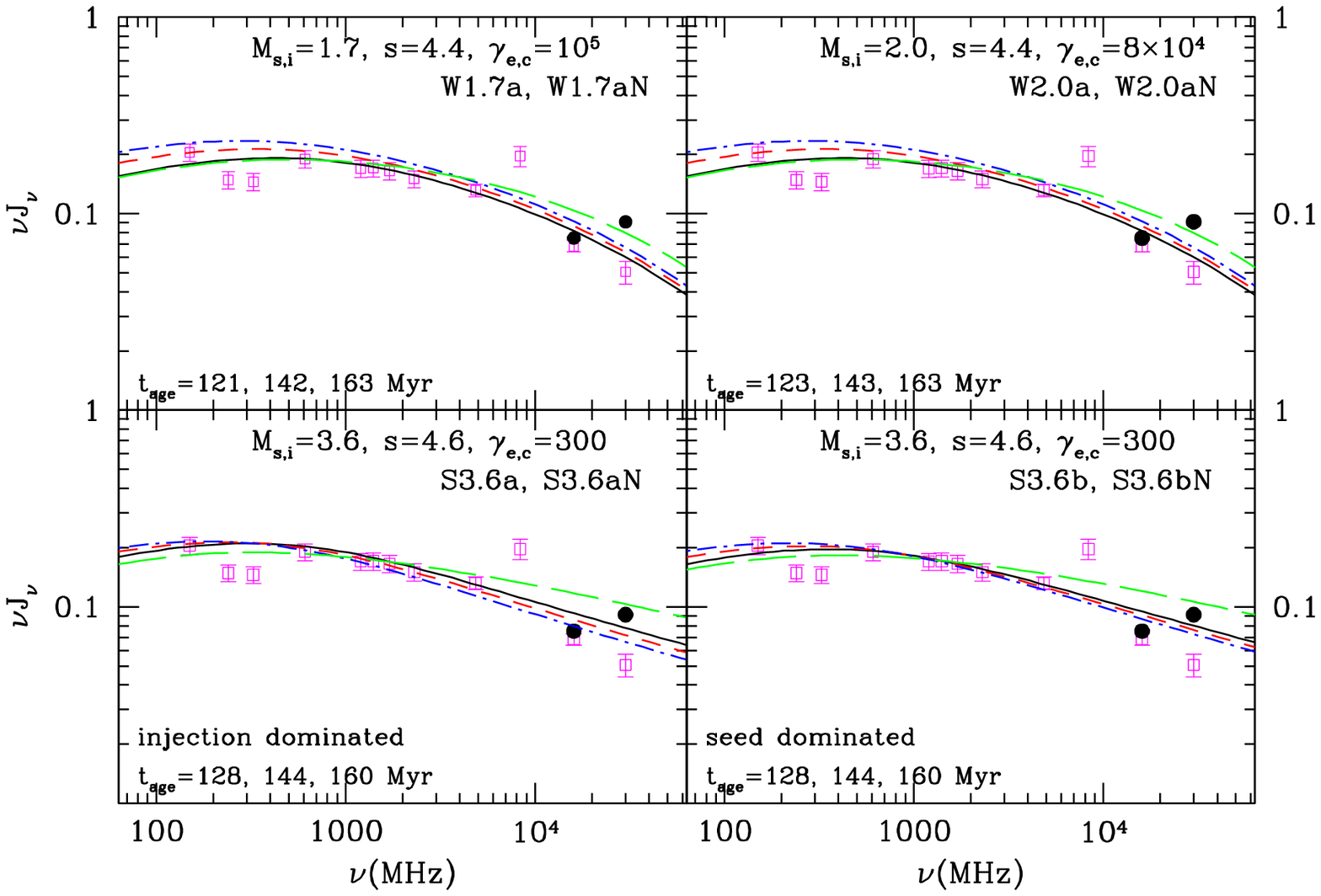}
\vspace{-6.3cm}
\caption{Time evolution of the volume-integrated synchrotron spectrum, $\nu J_{\nu}$, for the W1.7a, W2.0a, S3.6a, and S3.6b models.
See Table 1 for model parameters.
The spectra at three different shock ages are shown with black solid, red dashed, and blue dot-dashed lines.
The green long-dashed line shows $\nu J_{\nu}$ at the first epoch for models without TA.
Note that the normalization factors for the green lines are 1.6 times higher than for other models with TA.
The open magenta squares and solid black filled circles are for the B1 component of the Toothbrush relic.
The squares at low frequencies are the observational data given in Table A1 of \citet{stroe16}.
The two squares at 4.85 and 8.35 GHz are fluxes in Table 5 of \citet{kierdorf16}, multiplied by a factor of 0.71.
The error bars are given in the same tables.
The solid black circles at 16 and 30 GHz are the data points, multiplied by factors of 1.1 and 1.8, respectively, which could represent the SZ-corrected fluxes \citep{basu16}.
}
\label{Fig5}
\end{figure*}

In the upper panels of Figure 4, different normalization factors are adopted for each model to obtain the best match with 
the observed flux level of $S_{150}$ roughly at the peak values near $30-50$~kpc.
The same relative normalization factors are scaled for the higher frequency and applied to $S_{610}$ in the middle panels.
The observed profile of $S_{150}$ indicates that the region of the Toothbrush relic beyond $R>150$ kpc 
might be contaminated by a contribution from the radio halo.

We find that for the W1.7 and W2.0 models, a preexisting electron population with $s=4.4$ and 
$\gamma_{e,c}\ga 4\times 10^4$ is necessary to reproduce the observed spectral steepening profile across the relic width.
Moreover, the results demonstrate that the six models with TA (W1.7a,b, W2.0a,b, and S3.6a,b) 
can reproduce the observed profiles of $S_{\nu} (R) $ and $\alpha_{150}^{610} (R)$ reasonably well, 
while, as noted previously, none of the models without TA (black solid lines) can reproduce the profile of $\alpha_{150}^{610}$.
However, it is also important to realize that the models should not produce ``excess'' TA.
In particular, also as noted previously,
the W2.0d model (green) with $\tau_{\rm acc,0}=50$ Myr and 
the S3.6c model (green) without turbulence decay produce ``too-flat'' profiles of $\alpha_{150}^{610}$.

At the time of observation, $M_{\rm s,o}\approx 3.03$, in the S3.6 models, so $\alpha_{\rm s}\approx 0.74$,
which is slightly flatter than the observed index of 0.8 at the leading edge of the Toothbrush relic.
This, we argue, is still consistent, because the observed radio flux profiles are blended by a finite telescope beam. 
We also considered a model (not shown) with $M_{s,i}=3.3$ with $M_{\rm s,o}\approx 2.85$, so at the time of observation, $q \approx 4.6$ ($\alpha_{\rm s} \approx 0.78$). That model, however, produces a spectral index profile across the relic
a bit too steep to be compatible with the observed profile.

\subsection{Volume-Integrated CRe and Radio Spectra}

\begin{figure*}[t]
\vspace{-0.8cm}
\hspace{-0.2cm}
\includegraphics[scale=0.9]{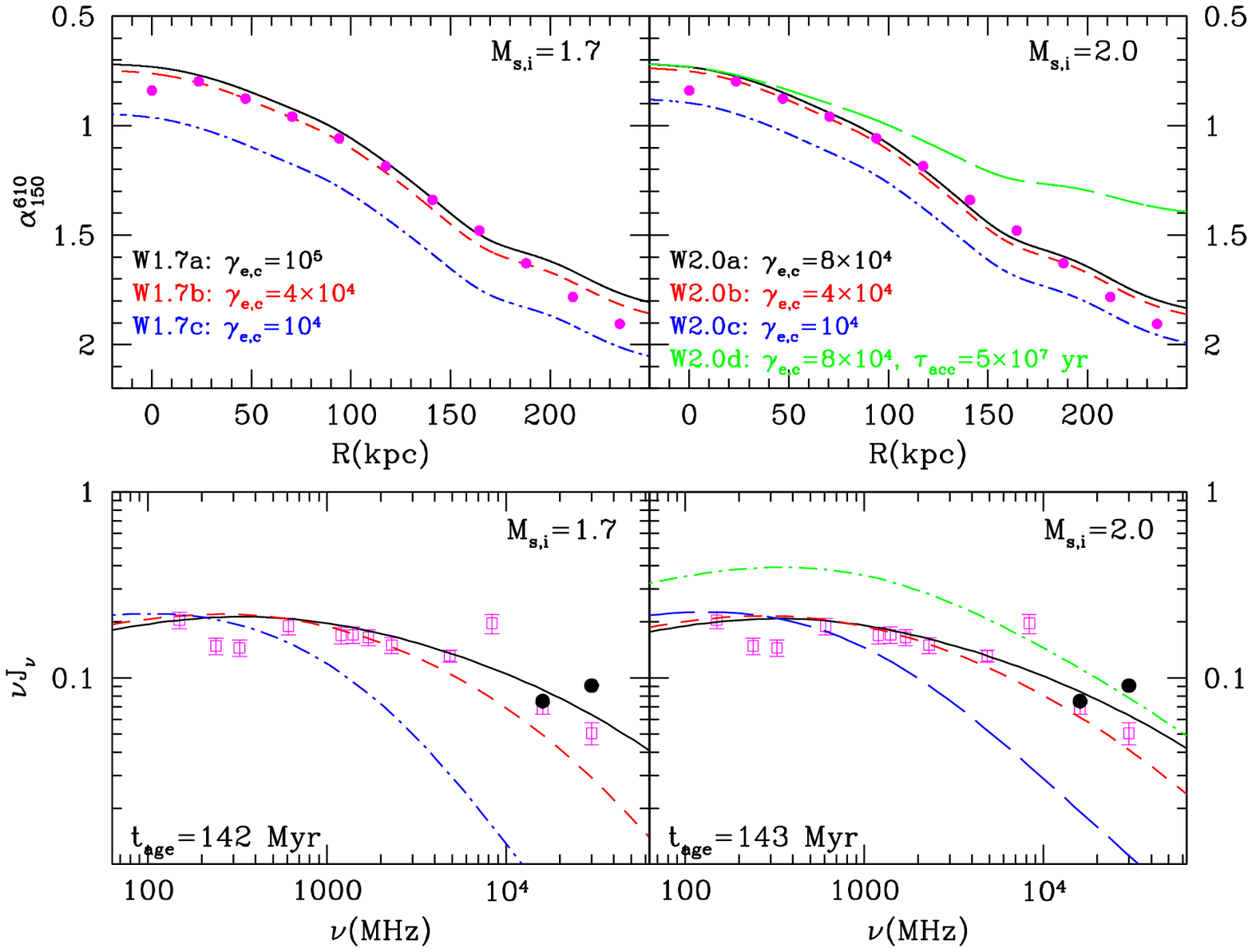}
\vspace{-5.3cm}
\caption{
Spectral index between 150 and 610 MHz, $\alpha_{150}^{610} (R)$, (top panels)
and volume-integrated synchrotron spectrum, $\nu J_{\nu}$, (bottom panels) for the weak-shock models.
The models with different values of $\gamma_{e,c}$ are compared (W1.7a,b,c and W2.0a,b,c).
In the W2.0d model with $\tau_{\rm acc}=5\times10$ Myr (green long-dashed lines),
turbulent acceleration is faster than in the W2.0a model.
The magenta dots in the upper panels are the same as those in Figure 4.
The open magenta squares and solid black filled circles in the lower panels are the same as those in Figure 5.}
\label{Fig6}
\end{figure*}

In the case of pure {\it in situ} injection without TA, 
the postshock momentum distribution function is basically the same as
the DSA power-law spectrum given in Equation (\ref{finj}) 
except for the increasingly lower exponential cutoff due to postshock radiative cooling.
So the volume-integrated CRe energy spectrum, $F_{\rm e}(p)=\int f_{\rm e}(r,p) dV$,
is expected to have a broken power law form, whose slope increases from $q$ to $q+1$ 
at the break momentum, $p_{\rm br}/m_{\rm e}c \approx 10^4 (t_{\rm age}/100 {\rm Myr})^{-1}(5\muG)^2/(B_2^2+B_{\rm rad}^2)$.
In the lower right panel of Figure 1, for instance, we can see that the volume-integrated electron spectrum, $G_{\rm e}(p)=p^4 F_{\rm e}(p)$, steepens gradually near $p/m_{\rm e}c \sim 3\times 10^3$
in the S3.6aN model (without TA, green long-dashed line).

Of course such a simple picture for the steepening does not apply to the W1.7 and W2.0 models with the DSA reacceleration
of preexisting electrons,
since the spectrum at the shock, $f_{\rm reacc}$, is a broken power-law that steepens from $p^{-s}$ to $p^{-q}$
above $p_{\rm e,c}$. In these models, $G_{\rm e}(p)$ 
depends on the assumed value of $\gamma_{e,c}$ (see the black, red, and blue lines in the lower left and lower middle panels of Figure 1) 
as well as $\tau_{\rm acc}$.
The models without TA are also shown as green long-dashed lines for comparison.

In the S3.6a model in the lower right panel of Figure 1, the suprathermal $\kappa$-like population for $p\ga p_{\rm inj}\approx 30\ m_{\rm e} c$ provides seed electrons 
for the {\it in situ} injection into DSA and subsequent TA in the postshock flow. 
In fact, this results in an excess, low-energy CRe population in the range
$30\la p/m_{\rm e}c \la300$ for the models, 
compared to the S3.6aN model, as shown in the figure.
This low-energy component depends on the details of kinetic plasma processes operating near the shock, which are not yet fully understood,
and would not contribute significantly to the observed radio emission in the range of $0.15-10$ GHz. 
For the postshock magnetic field strength, $B_2\approx 2.5 \muG$, electrons with $6.9 \times 10^3 \le p/m_{\rm e}c \le 5.6\times 10^4$
make the peak contribution in this observation frequency range.

From the spectral shape of $G_{\rm e}(p)$, we expect that the ensuing volume-integrated radio spectrum, $J_{\nu}=\int j_{\nu}(r)dV$, 
should steepen gradually toward high frequencies. Moreover, the form
depends on $p_{e,c}$ and $\tau_{\rm acc}$ in the W1.7 and W2.0 models and on $p_{\rm br}$
and $\tau_{\rm acc}$ in the S3.6 models.

Figure 5 shows the volume-integrated radio spectrum, $\nu J_{\nu}$, for the W1.7a, W2.0a, S3.6a, and S3.6b models
at three different shock ages to demonstrate how the spectrum evolves in time.
For the models without TA (W1.7aN, W2.0aN, S3.6aN, and S3.6bN), the spectrum is shown 
only at the first epoch (the green long-dashed lines). 
In each panel, the normalization factor for the vertical scale is chosen so that the simulated curves
match the observation data around 2 GHz.
For the models without TA, 
the normalization factor is 1.6 times larger than for the corresponding models with TA.
Note that the open squares (except at 4.85 and 8.35 GHz) are data for the B1 component of the Toothbrush relic in Table A1 of
\citet{stroe16}. \citet{kierdorf16} presented the sum of B1 + B2 + B3 flux at 4.8 and 8.35 GHz
in their Table 5.
Considering that the average ratio of the $B1/(B1+B2+B3)$ fluxes near 2 GHz is about 0.71
according to Tables 3 and A1 of \citet{stroe16}, we lower the fluxes at 4.85 and 8.35 GHz in the Kierdorf's data
by the same factor. 
\citet{basu16} showed that the Sunyaev-Zel’dovich (SZ) decrement in the observed radio flux can be significant above 10 GHz 
for radio relics.
We adopt their estimates for the SZ contamination factor for the Toothbrush relic given in their Table 1.
Then the SZ correction factors, $F$, for the fluxes at 16 and 30 GHz are about 1.1 and 1.8, respectively. 
Two solid black filled circles correspond to the flux levels so-corrected at the two highest frequencies.

Although the models without TA do not reproduce the observed profile of $\alpha_{150}^{610} (R)$, as shown in Figure 4,
the W1.7aN and W2.0aN models seem to fit the observed $J_{\nu}$ better than the W1.7a and W2.0a models.
So this exercise teaches us that it is important to test any model against several different observed properties.
Among the strong-shock models, S3.6a and S3.6b with TA
seem to produce better fits to SZ-uncorrected $J_{\nu}$, 
while S3.6aN and S3.6bN without TA give the spectra more consistent with SZ-corrected $J_{\nu}$.
In all models considered here, however, it seems challenging to explain the observed flux at 8.35 GHz.

In conclusion, adjustments of basic parameters can allow
both of the weak-shock and strong-shock models
to explain the observational data for the Toothbrush B1 component reasonably well.
In the weak-shock scenario, as we argued in the Introduction, however, it would be challenging to fulfill 
the requirement for a homogeneous, flat-spectrum preexisting electron population over a region 
400 kpc in length and $300$ kpc in width, which is needed to explain the observed uniformity 
in the spectral index along the length of the relic.
If the preexisting electrons cool by radiative and collisional losses non-uniformly, 
or if the preshock CRe have a span in ``ages'', both the cutoff energy and thus
the spectral index at the relic edge would be expected to vary along the relic length.

To explore such effects, we compare in Figure 6 the weak-shock models allowing different cutoff energies, 
$10^4 \le \gamma_{e,c} \le 10^5$. 
In order to reproduce the observed profiles of both $\alpha_{150}^{610}$ and $\nu J_{\nu}$,
$\gamma_{e,c}\ga 8\times 10^4$ is required for the W1.7 and W2.0 models.
Considering that the cooling times for electrons with $\gamma_{e,c} = 8\times 10^4$ in microgauss fields
are only $\sim 13$ Myr, it would be very challenging to explain a constant $\gamma_{e,c}$ within the required preshock region.

In the right-hand panels of Figure 6, the W2.0d model (green long-dashed lines) shows that
the ``enhanced'' TA with $\tau_{\rm acc,0}=50$ Myr would be too efficient to explain
the observed profile of $\alpha_{150}^{610} (R)$. The model produces too many low-energy electrons
with $\gamma_{\rm e}<10^4$, compared to high-energy electrons with $\gamma_{\rm e}\ga 10^4$.
This implies that the path to a model consistent with the observations 
cannot involve the adoption of smaller $\gamma_{e,c}$ combined with more rapid TA (smaller $\tau_{\rm acc}$).

Our results indicate that the strong-shock model with $M_{\rm s}\approx 3$ is favored.
That could mean that the observed X-ray and radio Mach numbers represent different parts of a nonuniform shock surface (see the discussion in the Introduction).
However, we should point out that the predicted $J_{\nu}$ values for the S3.6a and S3.6b models deviate from the observed
curvature at 8.35 GHz (Figure 5). 
Finally, as noted earlier, in order to explain the rareness of detected radio relics in merging clusters,
radio relics might be generated preferentially when shocks encounter regions of preexisting low-energy CRe (i.e., the S3.6b model).

\section{SUMMARY}

In this study, we reexamine the merger-driven shock model for radio relics, in which relativistic electrons 
are accelerated via DSA at the periphery of galaxy clusters.
To that end, we perform time-dependent DSA simulations of one-dimensional, spherical
shocks, and we compare the results with observed features of the Toothbrush relic
reported by \citet{stroe16} and \citet{vanweeren16}.
In addition to DSA, energy losses by Coulomb scattering, synchrotron emission, and iC scattering off the 
CMB radiation, and, {\it significantly}, TA by compressive MHD/plasma mode downstream of the shock
are included in the simulations.

Considering apparently incompatible shock Mach numbers from X-ray ($M_{\rm X}\approx 1.2-1.5$) and radio ($M_{\rm rad}\approx 2.8$) observations of the Toothbrush relic,
two possible scenarios are considered (see Table 1 for details):
(1) {\it weak-shock} models in which a preexisting flat-spectrum electron population with high cutoff energy is accelerated 
by a weak shock with $M_{\rm s}\approx 1.6-1.9$,
and (2) {\it strong-shock} models in which low-energy seed CRe, either shock-generated suprathermal electrons 
or preexisting soft-spectrum electrons, are accelerated by a strong shock with $M_{\rm s}\approx 3.0$.

The main results are summarized as follows:

1. In order to reproduce the broad profile of the spectral index behind the head (component B1) of the Toothbrush relic, 
TA with $\tau_{\rm acc}\approx 100$ Myr should be included to delay the spectral aging in the postshock region. This level of TA is strong but plausible in ICM postshock flows.

2. The strong-shock models with $M_{\rm s}\approx 3.0$, either with a
$\kappa$-like distribution of suprathermal electrons (the S3.6a model) or with low-energy preexisting CRe
with $p/m_{\rm e}c\la 300$ (the S3.6b model), are more feasible than the weak-shock models. 
These models could explain the observed uniform spectral index profile along
the relic edge over 400 kpc in relic length (component B1).
Further, the S3.6b model may be preferred
because (1) it can reproduce the observed flux density with a small fraction ($N\approx 0.05\%$) of preexisting CRe,
and (2) it can explain the low occurrence ($\la 10 \%$) of giant radio relics among merging clusters, 
where otherwise ``suitable'' shocks are expected to be common.
These low-energy fossil electrons could represent the leftovers either previously accelerated within the ICM by 
shock or turbulence or ejected from AGNs into the ICM, since their cooling times are long, $t_{\rm rad}> 3.5$~Gyr 
with $B\sim 1 \muG$ for $\gamma_{e} <300$.
The the S3.6a model, in which a $\kappa=1.6$ suprathermal distribution is adopted, the predicted flux density is about
five times smaller than the observed level.

3. For the weak-shock models with $M_{\rm s}\approx 1.6-1.9$,
a flat ($s\approx 4.4$) preexisting electron population with seemingly unrealistically high-energy cutoff 
($\gamma_{e,c}\ga 8\times 10^4$) is required to reproduce the observational data (the W1.7a and W2.0a models).
It would be challenging to generate and maintain such a flat-spectrum preexisting population with a
uniform value of $\gamma_{e,c}$ over the upstream region of 400 kpc in length and $300$ kpc in width,
since the cooling time is short, $\tau_{\rm rad} \sim 10$~Myr
for electrons with $\gamma_{\rm e} \sim 10^5$ in a $1\muG$ level magnetic field.

\acknowledgements
H.K. was supported by the Basic Science Research Program through the National Research Foundation of Korea (NRF) funded by the Ministry of Education (2014R1A1A2057940).
D.R. was supported by the National Research Foundation of Korea through grants 2014M1A7A1A03029872 and 2016R1A5A1013277.
T.W.J. was supported by the US National Science Foundation through grant AST1211595.
The authors thank R. J. van Weeren for providing the radio flux data for the Toothbrush relic
published in \citet{vanweeren16}.

\software{CRASH \citep{kj06}}


\begin{thebibliography}{}

\bibitem[Akamatsu \& Kawahara(2013)]{akamatsu13}
Akamatsu, H. \& Kawahara, H. 2013, 
\pasj, 65, 16

\bibitem[Basu et al.(2016)]{basu16}
Basu, K., Vazza, F., Erler, J., \& Sommer, M. 2016,
\aap, 591, A142

\bibitem[Br\"uggen et al.(2012)]{brug12} 
Br\"uggen, M., Bykov, A., Ryu, D., \& R\"ottgering, H. 2012, 
\ssr, 166, 187

\bibitem[Brunetti \& Jones(2014)]{brunetti2014}
Brunetti, G. \& Jones, T. W. 2014, 
IJMPD, 23, 30007

\bibitem[Brunetti \& Lazarian(2007)]{brunetti2007}
Brunetti, G. \& Lazarian, A. 2007, 
\mnras, 378, 245

\bibitem[Brunetti \& Lazarian(2011)]{brunetti2011}
Brunetti, G. \& Lazarian, A. 2007, 
\mnras, 412, 817

\bibitem[Brunetii \& Lazarian(2016)]{brunetti2016}
{\bf Brunetti, G. \& Lazarian, A. 2016,
\mnras, 458, 2584}

\bibitem[Caprioli et al.(2015)]{caprioli15}
Caprioli, D., Pop, A. R., \& Sptikovsky, A. 2015,
ApJ, 798, 28

\bibitem[Caprioli \& Spitkovsky(2014)]{caprioli14}
Caprioli, D. \& Sptikovsky, A. 2014,
ApJ, 783, 91

\bibitem[Donnert et al(2016)]{donnert16}
Donnert, J. M. F., Stroe, A., Brunetti, G., et al. 2016,
\mnras, 462, 2014

\bibitem[Drury(1983)]{dru83}
Drury, L. O'C. 1983,
RPPh, 46, 973

\bibitem[En{\ss}lin(1999)]{ensslin99}
En{\ss}lin, T. A. 1999, in Ringberg Workshop on Diffuse Thermal
and Relativistic Plasma in Galaxy Clusters, ed. P. S. H.
B\"ohringer, L. Feretti, vol. 271 of MPE Report, 275

\bibitem[En{\ss}lin et al.(1998)]{ensslin98}
En{\ss}lin, T. A., Biermann, P. L., Klein, U., \& Kohle S. 1998,
\aap, 332, 395

\bibitem[En{\ss}lin \& Gopal-Krishna(2001)]{ensslin01}
En{\ss}lin, T. A. \& Gopal-Krishna, 2001,
\aap, 366, 26

\bibitem[Feretti et al.(2012)]{feretti12}
Feretti, L., Giovannini, G., Govoni, F., \& Murgia, M. 2012,
A\&A Rev., 20, 54

\bibitem[Fujita et al.(2015)]{fujita2015}
Fujita, Y., Takizawa, M., Yamazaki, R., Akamatsu, H., \& Ohno, H., 2015
\apj, 815,116

\bibitem[Gieseler et al.(2000)]{gieseler2000}
Gieseler, U. D. J., Jones T. W., \& Kang, H. 2000, \aap, 364, 911

\bibitem[Guo et al.(2014)]{guo14}
Guo, X., Sironi, L., \& Narayan, R. 2014,
\apj, 793, 153

\bibitem[Heavens \& Meisenheimer(1987)]{heavens87}
Heavens, A. F. \& Meisenheimer, K. 1987, \mnras, 225, 335

\bibitem[Hong et al.(2015)]{hong15}
Hong, E. W., Kang, H., \& Ryu, D. 2015,
\apj, 812, 49

\bibitem[Hong et al.(2014)]{hong14}
Hong, E. W., Ryu, D., Kang, H., \& Cen, R. 2014,
\apj, 785, 133

\bibitem[Kang(2011)]{kang11}
Kang, H. 2011,
JKAS, 44, 49

\bibitem[Kang(2015)]{kang15}
Kang, H. 2015,
JKAS, 48, 155

\bibitem[Kang(2016a)]{kang16a}
Kang, H. 2016a,
JKAS, 49, 83

\bibitem[Kang(2016b)]{kang16b}
Kang, H. 2016b,
JKAS, 49, 145

\bibitem[Kang \& Jones(2006)]{kj06}
Kang, H. \& Jones, T. W. 2006,
APh, 25, 246

\bibitem[Kang et al.(2002)]{kjg02}
Kang, H., Jones, T. W., \& Gieseler, U. D. J. 2002,
\apj, 579, 337

\bibitem[Kang \& Ryu(2015)]{kangryu15}
Kang, H. \& Ryu, D. 2015,
\apj, 809, 186

\bibitem[Kang et al.(2012)]{kang12}
Kang, H., Ryu, D., \& Jones, T. W. 2012,
\apj, 756, 97

\bibitem[Kang et al.(2014)]{kang14}
Kang, H., Vahe, P., Ryu, D., \& Jones, T. W. 2014,
\apj, 788, 141

\bibitem[Kierdorf et al.(2016)]{kierdorf16}
Kierdorf, M., Beck, R., Hoeft, M., et al. 2016,
\aap, 600, A18

\bibitem[Kowal \& Lazarian(2010)]{kowal2010}
Kowal, G. \& Lazarian, A. 2010,
\apj, 720, 742

\bibitem[Lynn et al.(2014)]{lynn2014}
Lynn, J. W., Quataert, E., Chandran, B. \& Parrish, I. J. 2014,
\apj, 791, 71

\bibitem[Miniati(2015)]{miniati2015}
Miniati, 2015,
\apj, 800, 60

\bibitem[Ogrean et al.(2014)]{ogrean14}
Ogrean, G. A., Br{\"u}ggen, M., van Weeren, R., et al. 2014,
\mnras, 440, 3416

\bibitem[Park et al.(2015)]{park15}
Park, J., Caprioli, D., \& Sptikovsky, A. 2015,
\prl, 114, 085003

\bibitem[Pfrommer et al.(2006)]{pfrommer2006} 
Pfrommer, C., Springel, V., En{\ss}lin, T. A., \& Jubelgas, M. 2006,
\mnras, 367, 113 

\bibitem[Pierrard \& Lazar(2010)]{pierrard10}
Pierrard, V. \& Lazar, M. 2010, SoPh, 265, 153

\bibitem[Pinzke et al.(2013)]{pinzke13}
Pinzke, A., Oh, S. P., \& Pfrommer, C. 2013,
\mnras, 435, 1061 

\bibitem[Porter et al.(2015)]{porter2015}
Porter, D. H., Jones, T. W., \& Ryu, D. 2015, \apj, 810, 93

\bibitem[Ptuskin(1988)]{ptuskin1988}
Ptuskin, V. S. 1988,
Sov. Astron. Lett., 14, 255

\bibitem[Ryu et al.(2003)]{ryu03} 
Ryu, D., Kang, H., Hallman, E., \& Jones, T. W. 2003, 
\apj, 593, 599

\bibitem[Ryu \& Vishniac(1991)]{ryu91} 
Ryu, D. \& Vishniac, E. T. 1991,
\apj, 368, 411

\bibitem[Sarazin(1999)]{sarazin99}
Sarazin C. L. 1999, 
\apj, 520, 529

\bibitem[Schekochihin \& Crowley(2006)]{schek2006}
Schekochihin, A. \& Crowley, S. 2006, PhPl, 13, 56501

\bibitem[Skilling(1975)]{skill75}
Skilling, J. 1975,
\mnras, 172, 557

\bibitem[Skillman et al.(2008)]{skillman2008} 
Skillman, S. W., O'Shea, B. W., Hallman, E. J., Burns, J. O., \& Norman, M. L. 2008, \apj, 689, 1063 

\bibitem[Stroe et al.(2016)]{stroe16}
Stroe, A., Shimwell, T. W., Rumsey, et al. 2016,
MNRAS, 455, 2402

\bibitem[Sunberg et al.(2016)]{sundberg16}
Sundberg, T., Haynes, C. T., Burgess, D., \& Mazelle, C., X. 2016,
\apj, 820, 21

\bibitem[van Weeren et al.(2016)]{vanweeren16} 
van Weeren, R. J., Brunetti, G., Br\"uggen, M., et al. 2016 
\apj, 818, 204


\bibitem[van Weeren et al.(2010)]{vanweeren10} 
van Weeren, R. R., R\"ottgering, H. J. A., Br\"uggen, M., \& Hoeft, M. 2010, 
Science, 330, 347

\bibitem[van Weeren et al.(2012)]{vanweeren12} 
van Weeren, R. J., R\"ottgering, H. J. A., Intema, H. T., et al. 2012, 
\aap, 546, A124 

\bibitem[Vazza et al.(2009)]{vazza09} 
Vazza, F., Brunetti, G., \& Gheller, C. 2009, 
\mnras, 395, 1333

\bibitem[Yan \& Lazarian(2002)]{yan2002}
Yan, H. \& Lazarian, A. 2002,
\prl, 89, 281102

\end{thebibliography}
\end{document}